\documentclass[journal=jacsat,manuscript=article]{achemso}

\usepackage[version=3]{mhchem} 
\usepackage{amsmath}
\usepackage{mathtools}
\usepackage{hyperref}
\usepackage{graphicx}
\usepackage{txfonts}
\usepackage{epstopdf}
\usepackage{natbib}
\usepackage{adjustbox}

\usepackage{nameref}
\usepackage{varioref}
\usepackage{hyperref}
\usepackage{cleveref}

\usepackage{xcolor}
\usepackage[normalem]{ulem}

\newif \ifreview
\reviewfalse      

\ifreview
  \usepackage{changes}

  \definecolor{ACSblue}{RGB}{0,60,125}
  \definecolor{ACSred}{RGB}{150,0,0}
  \definecolor{ACSpurple}{RGB}{90,0,110}

  \definechangesauthor[name={VS}, color=red]{VS}
  \definechangesauthor[name={FD}, color=blue]{FD}
  \definechangesauthor[name={BH}, color=green]{BH}
\else
  \usepackage[final]{changes}
\fi

\setaddedmarkup{\textcolor{ACSblue}{\uline{#1}}}

\setdeletedmarkup{\textcolor{ACSred}{\sout{#1}}}

\setcommentmarkup{\textcolor{ACSpurple}{\textbf{[#1]}}}

\author{Valeriia Sokolova}
\email{valeriia.sokolova@cyu.fr}
\affiliation[LIRA]
{CY Cergy Paris Université, Observatoire de Paris, Université PSL, Sorbonne Université, Université Paris Cité, CNRS, LIRA, F-95000, Cergy, France}
\author{Basile Husquinet}
\affiliation[MPE]
{Center for Astrochemical Studies, Max-Planck-Institute for Extraterrestrial Physics (MPE), Gie\ss enbachstr. 1, D-85748 Garching, Germany}
\alsoaffiliation[LIRA]
{CY Cergy Paris Université, Observatoire de Paris, Université PSL, Sorbonne Université, Université Paris Cité, CNRS, LIRA, F-95000, Cergy, France}
\author{Stephan Diana}
\affiliation[LIRA]
{CY Cergy Paris Université, Observatoire de Paris, Université PSL, Sorbonne Université, Université Paris Cité, CNRS, LIRA, F-95000, Cergy, France}
\author{Paola Caselli}
\affiliation[MPE]
{Center for Astrochemical Studies, Max-Planck-Institute for Extraterrestrial Physics (MPE), Gie\ss enbachstr. 1, D-85748 Garching, Germany}
\author{Emanuele Congiu}
\affiliation[LIRA]
{CY Cergy Paris Université, Observatoire de Paris, Université PSL, Sorbonne Université, Université Paris Cité, CNRS, LIRA, F-95000, Cergy, France}
\author{Wiebke Riedel}
\affiliation[MPE]
{Center for Astrochemical Studies, Max-Planck-Institute for Extraterrestrial Physics (MPE), Gie\ss enbachstr. 1, D-85748 Garching, Germany}
\author{Olli Sipil\"{a}}
\affiliation[MPE]
{Center for Astrochemical Studies, Max-Planck-Institute for Extraterrestrial Physics (MPE), Gie\ss enbachstr. 1, D-85748 Garching, Germany}
\author{Anton Vasyunin}
\affiliation[URFU]
{Research Laboratory for Astrochemistry, Ural Federal University, Kuibysheva St. 48 Yekaterinburg 620026, Russia}
\author{Valentine Wakelam}
\affiliation[BD]
{Laboratoire d’astrophysique de Bordeaux, Univ. Bordeaux, CNRS, B18N, allée Geoffroy Saint-Hilaire, 33615 Pessac, France}
\author{Francois Dulieu}
\affiliation[LIRA]
{CY Cergy Paris Université, Observatoire de Paris, Université PSL, Sorbonne Université, Université Paris Cité, CNRS, LIRA, F-95000, Cergy, France}
\email{francois.dulieu@cyu.fr}

\title[]
  {Using astrochemical models to simulate reactivity experiments on cold surfaces}

\abbreviations{ISM,COM,IR,RE,QMS,TPD,\replaced[id=VS]{RSS}{TSD}}

\begin{document}







\begin{abstract}
The development of molecular complexity during stellar and planetary formation owes much to the interaction of gas and dust. When the first astrochemical models including solid-state chemistry were developed more than forty years ago, data from dedicated laboratory experiments were limited. Since then, many groups have developed specific experimental set-ups to address this issue, but astrochemical models have rarely been directly confronted with these new results. We want to demonstrate \replaced[id=VS]{whether}{if} it is possible to use rate-equation-type astrochemical models developed in the context of the Interstellar Medium to compare them with laboratory astrophysics experiments.

In this work, we use the case of low-temperature hydrogenation of CO, which is known to lead to methanol, among other molecules. We carried out 9 experiments, varying the experimental parameters such as temperature and dose. We give quantitative results and take care \replaced[id=VS]{of detailing}{to detail} the vocabulary used in the experiments. We use astrochemical codes, \replaced[id=VS]{NAUTILUS}{Nautilus}, pyRate and MONACO, to reproduce our experimental conditions, which requires good control of the change of vocabulary and scales, especially for fluxes and time scales.

This work demonstrates that it is possible to use different astrochemical codes to compare \added[id=VS]{modelling results} directly with the output of experiments. There are discrepancies between models and experiments, as well as between models, but a fair agreement is achieved. We discuss the possible origin of the differences, which could originate from the chemical network or the difference in the description of physical processes.

\end{abstract}

\section*{Keywords}
Astrochemistry, Molecular processes, Laboratory astrophysics, Numerical modelling, ISM: molecules \comment[id=VS]{VS: Keywords have been added this way to overcome the restrictions of technical submission copy mode.}

\section{Introduction}\label{sec:intro}

Dust particles play an important role in the astrophysics of the interstellar medium, from thermodynamics and chemical processes to the dynamics of star formation. The chemical composition of dust particles depends on the initial chemical composition of the environment where dust particles were formed: there are silicate and graphite particles, particles consisting of silicon carbide, etc~\cite{tiel2005}. 

Within the dense regions of the interstellar medium (ISM), such as molecular clouds, cores, and denser objects, where the temperatures are low and \added[id=VS]{the} number densities of particles in gas are high, species adsorb to the surface of dust grains and can easily form molecular ices. The composition of ice mantles is complex and includes significant amounts of carbon monoxide (CO), carbon dioxide (\ce{CO2}), water (\ce{H2O}), methanol (\ce{CH3OH}), ammonia (\ce{NH3}) and possibly complex molecules \cite{McClure2023,2024A&A...683A.124R,2024A&A...686A..71N}. These ices are formed under very cold conditions, typically below 100~K and even below \replaced[id=VS]{10~K}{10K}, and are influenced by cosmic rays, UV radiation, and the processes of accretion and desorption. Studying \added[id=VS]{the} dust ice mantles is one of the important questions in modern laboratory astrochemistry: understanding the composition and behaviour of ice is pivotal for further exploration of processes that govern the chemical and physical evolution of the ISM. 

In recent years, many attempts have been made to understand how complex organic molecules and astrochemical ices can be formed on the surfaces of ice and dust by combining observations~\cite{2017MNRAS.469.2230M, 2017A&A...605L...3C, 2024A&A...684A.189V, 2025A&A...693A.222T}, chemical modeling~\cite{2017ApJ...842...33V, 0004-637X-862-2-140, 10.1093/mnras/stad3896, sipila2015benchmarking, 2016MNRAS.459.3756R, 2025A&A...695A.247J}, and laboratory experiments~\cite{2015MNRAS.448.1288F, 2016MNRAS.455.1702C, 2021A&A...652A..29K, Nakibov_2025, Ozhiganov_2024, 2017ApJ...849...80C}. Laboratory experiments include \added[id=VS]{the} experimental studies as well as \added[id=VS]{the} quantum-chemical and molecular dynamics calculations~\cite{2018IAUS..332....3V}. This approach can be considered as a bridge between observations and the theoretical part of studies, providing data derived from controlled experiments on special setups for both fields. Such setups are capable of recreating conditions of \added[id=VS]{the} ISM and allow scientists to simulate and study complex chemical processes that can happen in space. One of the main questions that laboratory astrochemistry can answer is what type of molecules we expect to see and actually detect in space~\cite{schlemmer2015laboratory}. From experimental studies, we can obtain data about compounds that can be found in spectroscopy in the range from UV to mm wavelengths and compare it with data that is provided from observations. In addition to that, a lot of information about chemical reactions can be derived: for example, binding energies of compounds (e.g. \citealt{Perrero2024}) and activation energies of reactions, reaction rates etc. 

Theoretical studies of the chemical evolution of the ISM include computational codes that implement various mathematical methods for calculating reaction rates, atomic and molecular abundances, etc. All astrochemical codes use chemical reaction databases that are based on the results of laboratory studies. Currently, the most common databases are UMIST 2013 (for gas-phase reactions)~\cite{2013A&A...550A..36M, 2024A&A...682A.109M} and KIDA (which takes into account not only gas-phase reactions but also reactions on the surface of dust particles and their ice mantles)~\cite{2012ApJS..199...21W, 2015ApJS..217...20W,Wakelam2024}.

In this work, we focus on the study of the solid-state chemical evolution of atoms and molecules in the \{CO+H\} chemical system by a combination of laboratory experiments and astrochemical modeling. The hydrogenation of CO on the surface of dust particles is one of the most well-studied processes in astrochemistry and can be considered as a benchmark for chemistry on grain surfaces, thanks to its simplicity~\cite{2014A&A...572A..70R}. The hydrogenation of CO can eventually lead to the formation of formaldehyde and methanol through the sequence of reactions~\cite{Hiraoka2002, 2004ApJ...616..638W, fuchs, Pirim2011, Mini2016,2018sf2a.conf...15F}:
 \begin{equation*}
 \text{CO}  \xleftrightarrow[  ]{H} \text{HCO} \xleftrightarrow[  ]{H} \text{H}_{2}\text{CO} \xleftrightarrow[  ]{H} \text{CH}_{2}\text{OH} /\text{CH}_{3}\text{O} \xleftrightarrow[  ]{H} \text{CH}_{3}\text{OH},
 \end{equation*}
where forwards and backwards arrows indicate the H-addition and H-abstraction reaction, respectively. \comment[id=VS]{VS: removed the tabbed 'where'.}These reactions were carefully studied using a combination of laboratory experiments and theoretical approaches such as quantum chemical calculations and simple chemical modeling. However, despite all efforts, our understanding of this process is still not complete. In \added[id=VS]{the} past years it was discovered that the standard scenario of CH$_3$OH formation is missing some of the ''reverse'' reactions for HCO and H$_2$CO \cite{Mini2016}, leading not only to the new destruction paths, but also to \added[id=VS]{the} unexpected effect of non-thermal desorption of compounds~\cite{2020A&A...634A.103D}, except for HCO~\cite{2020ApJ...897...56P}, \replaced[id=VS]{which}{that} adds some constraints to the hydrogenation path. In following years, each step of the chain was revised, adding more and more information about activation energies, branching ratios and other important parameters for \added[id=VS]{the} reaction rate calculations~\cite{MORISSET20191}. From the theoretical point of view, the research was focused both on quantum calculations and classical modelling of CO chemical evolution. \citealt{2014A&A...572A..70R} were the first to combine a quantum approach (based on the density functional theory (DFT) approach), deriving energy barriers and transition frequencies, with macroscopic astrochemical modelling (based on \added[id=VS]{the} rate equations approach), where they implemented this data. Later on \citealt{2021A&A...655A...9E, 2022ApJS..259...39E} studied the activation energy barriers of radical-radical interactions including the  \{CO+H\} system. In following studies, the research was focused on using \added[id=VS]{the} rate equations approach, studying new mechanisms involved in the chain (e.g.\added[id=VS]{,} non-diffusive mechanisms~\cite{2020ApJS..249...26J}), applying all theoretical data to \added[id=VS]{the} modelling of the ISM objects of interest \cite{2022A&A...657A..10S, 2021ApJ...917...44J, 10.1093/mnras/stac3449}.

\replaced[id=VS]{This paper is organized as follows: Section~\nameref{sec:experiments} is dedicated to laboratory measurements and description of experiments, Section~\nameref{sec:re} describes the theoretical approach that we used, in Sections~\nameref{sec:monaco}, ~\nameref{sec:nautilus} and \nameref{sec:pyrate} we present the results of astrochemical modelling using different popular codes, and in the Section~\nameref{sec:comparison} we summarize all the results.}{This paper is organized as follows: Section~\nameref{sec:experiments} is dedicated to laboratory part of the study, covering conditions and description of all experiments, Section~\nameref{sec:re} describes the theoretical approach that we used, in Sections~\nameref{sec:monaco}, ~\nameref{sec:nautilus} and \nameref{sec:pyrate} we present the results of astrochemical modelling using different popular codes, and in the Section~\nameref{sec:comparison} we summarize all results and analysis of this study.}

\section{Experimental methods, vocabulary and results}\label{sec:experiments}
For this project, we use the special laboratory setup that provides controlled conditions that mimic \replaced[id=VS]{those}{ones} in space~--- VENUS, from the acronym VErs des NoUvelles Synthèses (''Towards new syntheses'' in French). The full description of the setup with all details can be found in~\citealt{congiu2020new}. 

\begin{figure*}
\center    \includegraphics[width=0.8\textwidth]{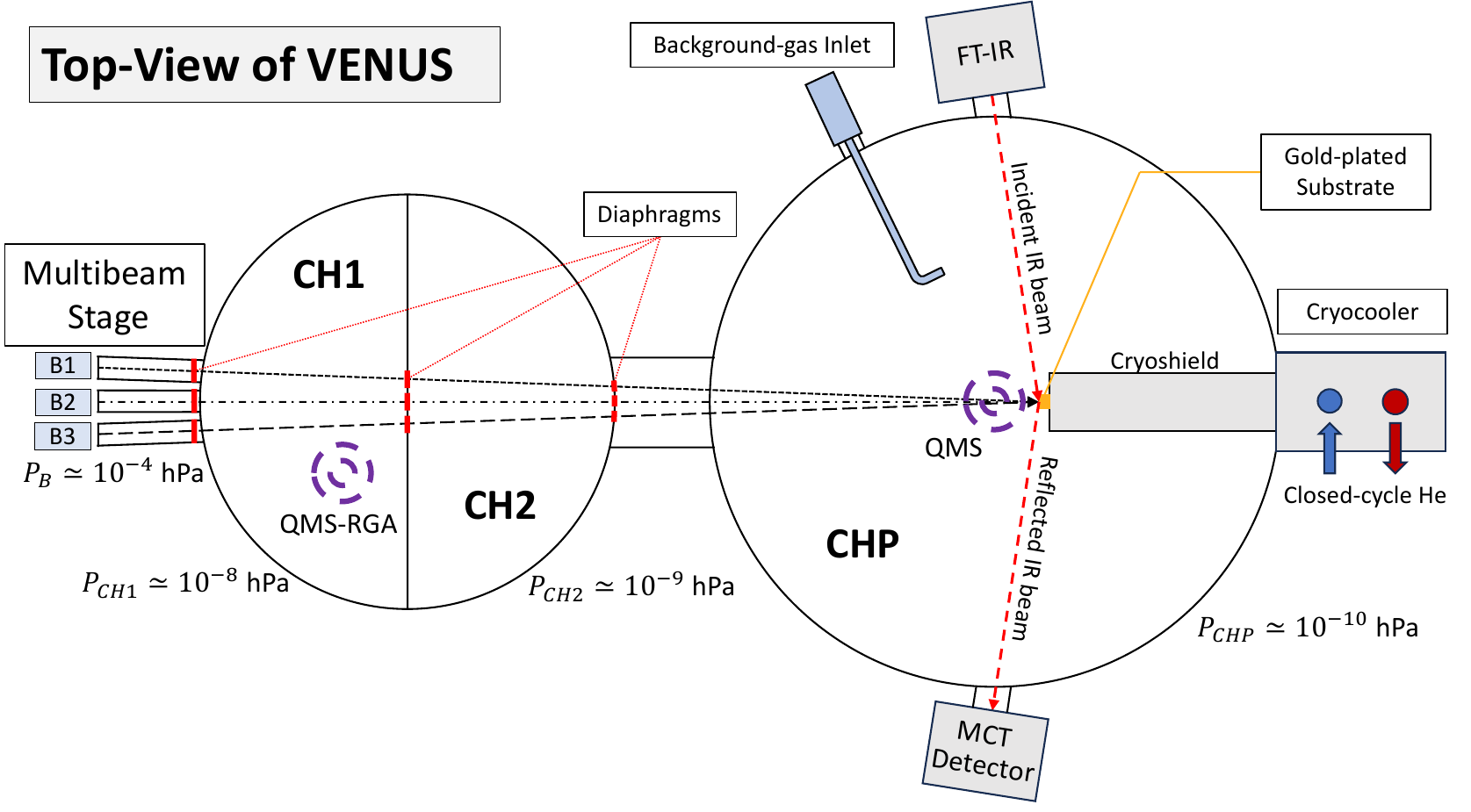}
    \caption{Simplified schematic top view of the VENUS Setup. From left to right: multibeam stage for species injection in the system, intermediate chamber (CH1 and CH2) for preparation of the beams, main chamber with gold-coated sample inside of it, QMS for detection of gas-phase species (place in front of the sample during the TDP), FT-IR detector for controlling the growth of the ices, and cryostat that regulates the temperature in the system. Source: \citealt{husquinet2025neon}. \added[id=VS]{Reprinted with permission from [Husquinet et al., A\&A, 703, A16 (2025) https://doi.org/10.1051/0004-6361/202554408] published under the terms of the Creative Commons Attribution License (https://creativecommons.org/licenses/by/4.0). Copyright 2025, Astronomy and Astrophysics.}}
    \label{fig:scheme}
\end{figure*}

The VENUS setup can provide very low temperatures for the experiments (5 - 8~K with the proper shielding) and ultra-high vacuum. The lowest pressure that can be obtained is 2$\times$10$^{-11}$ mbar, but the usual one that is used for the experiments is 10$^{-10}$ mbar, which corresponds to a gas density of $\simeq$ 3$\times$10$^{6}$ particles per cm$^3$, mostly composed of H$_2$, with traces of others UHV pollutants such as, by decreasing partial pressure CO, CO$_2$, H$_2$O, Ar, CH$_4$, and other hydrocarbons~\cite{congiu2020new}. These values are good enough to mimic conditions for the solid-state synthesis of molecules in the interstellar medium and allow surface-surface and gas-surface reactions to dominate over gas-gas collisions. For studying the formation and destruction of molecules a \replaced[id=VS]{quadrupole}{Quadrupole} mass-spectrometer (QMS) and an \replaced[id=VS]{infrared spectrometer}{Infrared spectroscope} (that implements vibrational fingerprint spectra in the mid-infrared domain 2.5 - 15 $\mu$m) were placed in the Main Chamber. The full scheme of the setup is presented in the Fig.~\ref{fig:scheme}.

The setup was designed to study very thin systems (from sub-monolayers to several monolayers) where molecules can react on the cold gold-coated surface with a typical collision rate of 10$^{12}$ particles cm$^{-2}$s$^{-1}$. All molecules are injected in the chamber in gaseous form, originally coming from liquids, gases or solids. 

Up to 4 species can be put in the system with the multi-beam stage, containing the right beam (B3), the top beam (B1), the central beam (B2) and the left beam (not in use at this moment), which is shown in the Fig.~\ref{fig:scheme}. These beam lines have typical accretion flux at the surface of $1.67 \times 10^{12}$ molecules cm$^{-2}$s$^{-1}$ that gives for example 1~ML of CO ice growing on the gold plate surface in approx. 10 minutes~\cite{congiu2020new}. Here and later all deposition doses will be quantified in monolayers (ML), and 1 ML  is 1$\times$10$^{15}$ molecules cm$^{-2}$ that corresponds to the number of adsorption sites per cm$^{-2}$ on the surface and has a physical meaning of one layer of product that covers the surface without desorbing (for example, CO at surface temperatures below 23~K). H atoms have low binding energy and high reactivity, so in that case an accretion of 1 ML of H does not mean that one layer of H atoms is present on the surface. Instead, we accept that statistically each adsorption site has been exposed to one H atom coming from the gas phase.

The general experiment consists of two steps: deposition of species on the surface (step 1) and sublimation of the evolved species in the gas during the linear increase of the surface temperature (step 2), usually \replaced[id=VS]{Temperature}{Thermally} Programmed Desorption (TPD). 

\begin{figure}
\center    \includegraphics[width=0.7\linewidth]{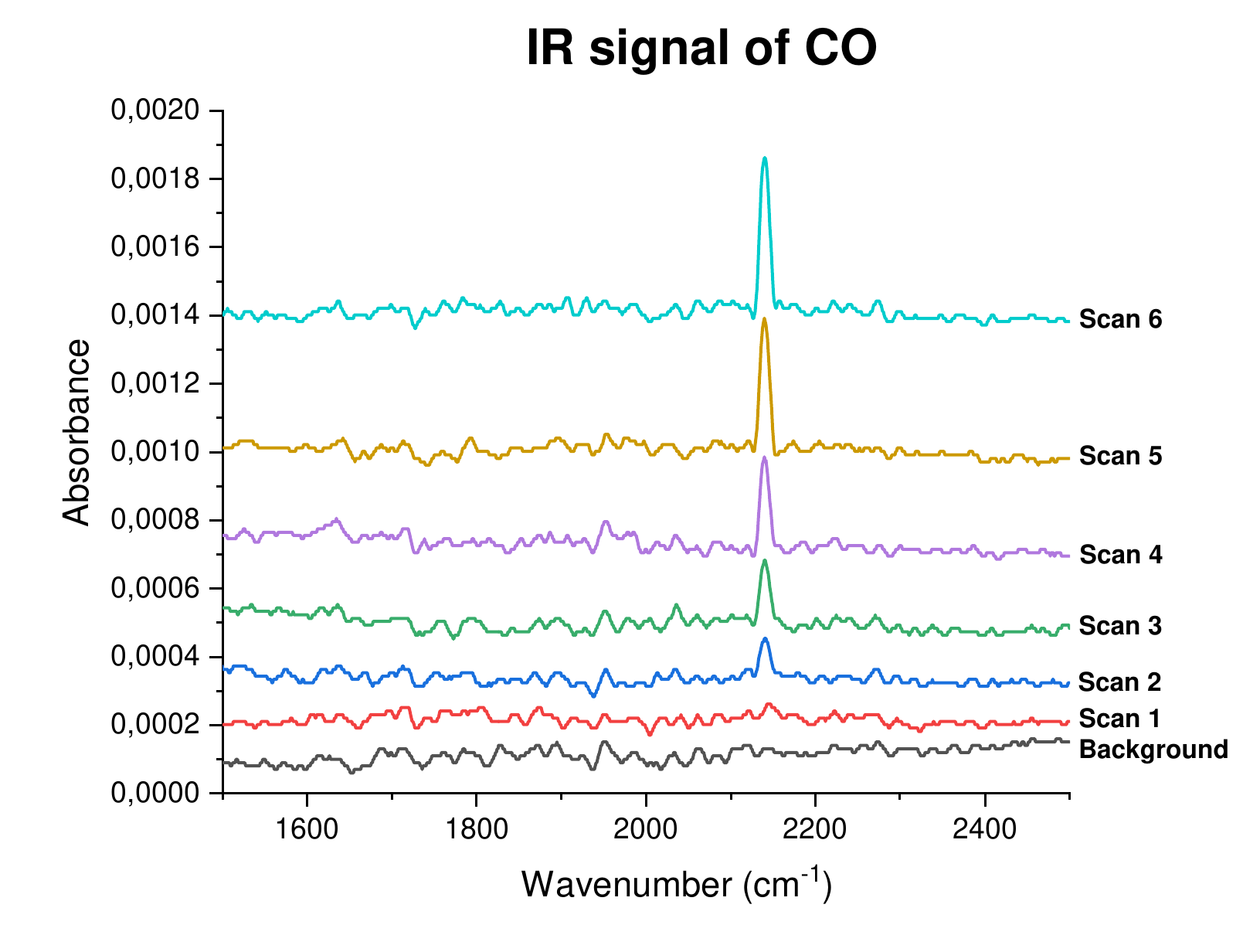}
    \caption{Family of infrared spectra for a CO deposition of 10 minutes on the gold surface held at 10 K (corresponding to experiment \# 1 of Table~\ref{tab:experiments} and \textit{A)} on Fig.~\ref{fig:reactions}). Scanning starts with the background spectrum and continues for the whole time of the deposition. Spectrum 6 refers to the additional scan made after the deposition was stopped, corresponding to 1.0 ML. A strong peak is visible at around 2140 cm$^{-1}$. Signals have been vertically offset for better visibility.}\label{fig:COMLSpec}
\end{figure}

\begin{figure}
\center    \includegraphics[width=0.7\linewidth]{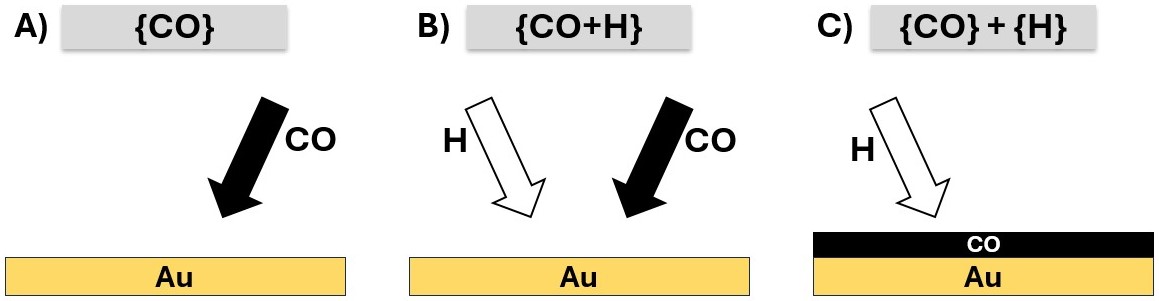}
    \caption{Schematic description of experiments shown in Table~\ref{tab:experiments}. CO and H are deposited on the gold-coated surface with a temperature of 10\,K. From left to right: CO deposition \added[id=VS]{(pic. \textit{A})}, co-deposition of H and CO \added[id=VS]{(pic. \textit{B})}, sequential deposition CO and H (or hydrogenation of CO ice layer, \added[id=VS]{pic. \textit{C}}).}\label{fig:reactions}
\end{figure}

The gold-coated copper sample in the main chamber is attached to the Helium cryostat's cold finger, which allows to keep a constant temperature during the whole time of the deposition. Deposited molecules come to the surface, and form the ice film, and during this step, they can be constantly monitored with Fourier transform reflection\added[id=VS]{-}absorption infrared spectroscopy (FT-RAIRS). This technique is used for analysing the chemical evolution from the moment of molecule deposition on the sample surface until the formation of new products and ices. Hence composition, thickness and morphology of ices can be studied ''on the go''. One saved infrared spectrum contains 256 averaged scans, corresponding to an acquisition time of approx. 2 min~\cite{congiu2020new}. Figure~\ref{fig:COMLSpec} presents the first spectra obtained during CO deposition on a surface held at 10 K for 10 min deposition on a bare gold surface (corresponding to experiment \# 1 of Table~\ref{tab:experiments} and to sketch \textit{A} of Figure~\ref{fig:reactions}). The column density of the adsorbate can be deduced from the area of the IR absorption peaks. In this case it corresponds to a linear quantity increase of CO on the surface.

The FT-RAIRS technique is combined with \added[id=VS]{the} TPD for better control of the evolution of chemical compounds on the sample surface. \added[id=VS]{The} TPD is a powerful tool for monitoring interactions of molecules on the surface during the controlled changing of the sample temperature. During the TPD, temperature is linearly increasing with a rate of 0.2~K~s$^{-1}$ and can be varied from 8 to 400\,K~\cite{congiu2020new}. All desorbing species are measured with the Hiden 51/3F vertically translatable quadrupole mass spectrometer which can monitor several masses at the same time and give the intensity of each chosen mass as a function of temperature. During this stage, QMS is placed at 5\,mm in front of the surface. This method provides information about the desorption temperatures (hence, indirect access to binding energies) of molecules and the quantity of these adsorbed molecules (from the intensity of \added[id=VS]{the} TPD spectrum peaks). 

\begin{figure}
    \centering
    \includegraphics[width=0.7\columnwidth]{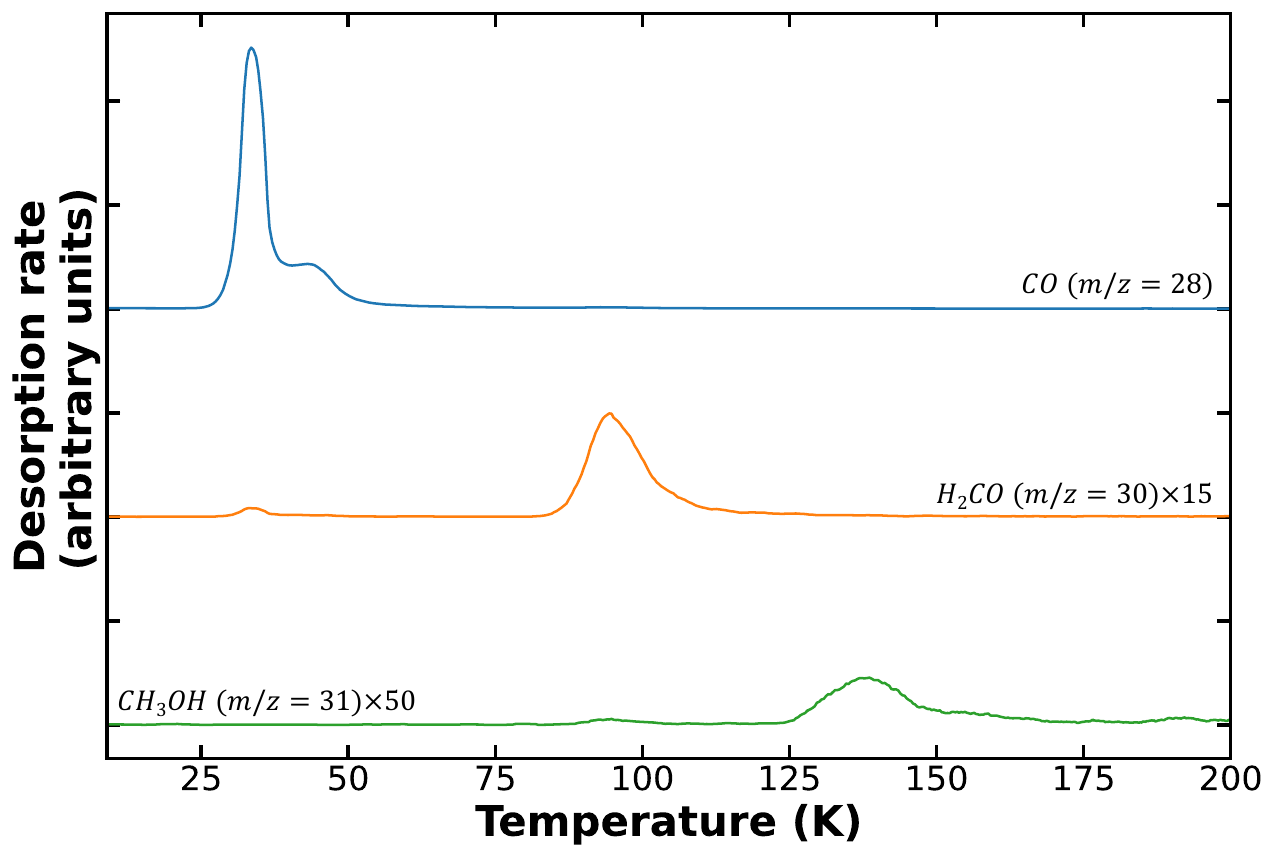}
    \caption{TPD spectra of the experiment \#4, a co-deposition of {CO+H} on a surface held at 9 K, where the top spectra in blue represent CO ($m/z$=28), \ce{H2CO} is represented in the orange spectra in the middle and the \ce{CH3OH} ($m/z$=31) in the bottom with the green spectra. Given the low levels of \ce{H2CO} and \ce{CH3OH} production, the spectra were multiplied by factors of $\times15$ and $\times50$, respectively.}
    \label{fig_exp: Example TPD}
\end{figure}

In our experiments, we varied the TPD from surface temperature (9---15\,K) to 200\,K with a ramp of 0.2\,K/s. Figure~\ref{fig_exp: Example TPD} shows the TPD spectrum of experiment \#4, displaying the main masses of CO ($m/z$=28), \ce{H2CO} ($m/z$=30), and \ce{CH3OH} ($m/z$=31). The abundances of each molecule ($N_x$) in Table 1 are calculated using the following equation:
\begin{equation}
    N_X = \frac{A_X}{A^{ML}_{CO}}\frac{\sigma^{Tot}_{CO}}{\sigma^{Tot}_{X}},
    \label{eq_exp: Quantities mol X}
\end{equation}
where $A_X$ represents the area of the TPD spectrum of molecule $X$, taking into account the cracking pattern (see \citealt{2024MNRAS.533...52V}). $A_{CO}^{ML}$ denotes the area of a CO monolayer, while $\sigma_X^{Tot}$ signifies the total cross section for electron impact ionization (at 30 eV, the value we use) for molecule X and $\sigma_{CO}^{Tot}$ for CO. The total cross-section for CO is $\sigma_{CO}^{Tot}=1.296\,\AA^2$ \cite{hwang1996new}, for \ce{H2CO} $\sigma_{\ce{H2CO}}^{Tot}=2.74\,\AA^2$ \cite{vinodkumar2011electron}, and for \ce{CH3OH} $\sigma_{\ce{CH3OH}}^{Tot}=3.263\,\AA^2$ \cite{kumar2019electron}. The sensitivity of the QMS by mass is not considered here, as the masses of the studied molecules are very similar (28 m/z to 32 m/z), which has a negligible impact on the final result. As \ce{H2CO} and \ce{CH3OH} are the products of CO, the estimated monolayer (ML) abundance of the products is calculated as a function of CO.

In addition, the TPD technique allows to calibrate the flux of the molecules. In the case of CO, the molecular beam calibration was performed as described in~\citealt{congiu2020new} and \citealt{2024A&A...686A.236K}. To calibrate the CO flux on our setup, we deposit CO on the surface for different time periods, corresponding to different doses, and then perform the TPD. From the TPD shapes the transition from \added[id=VS]{the} first to \added[id=VS]{the} second layer can be determined. For this calibration the surface temperature is fixed during depositions to guarantee the homogeneous distribution of molecules on the surface. For our experiments, we estimate that 1\,ML of CO ice is formed in 10 minutes using a flow rate of 0.03\,sccm (standard cubic centimetres per minute, quantifies the flow rate) in the source, and T$_{surf}$ = 10~K, which gives us a flux equal to :
\begin{equation}\label{flux}
\phi_{CO} = \frac{10^{15} \text{molecules cm}^{-2}}{600 \text{s}} \simeq 1.67 \times 10^{12} \text{molecules cm}^{-2}\text{s}^{-1}
\end{equation}

For the flux of H atoms, we dissociate H$_2$ by creating a plasma in the quartz tube, surrounded by the microwave cavity (see details in~\citealt{2024MNRAS.533...52V}). In that case, the absolute calibration uncertainty of flux is larger because it is based on H reactivity in multi-reaction systems that are supposed to be barrierless: for example, \{O$_2$+H\} or \{NO+H\}~\cite{2012ApJ...750L..12C}. In our settings, one ML of H atoms can be achieved in about 2.5 minutes (time to consume O$_2$ molecules), with flux $\phi_{H}$ = 6.67$ \times 10^{12} \text{atoms cm}^{-2}\text{s}^{-1}$ as can be seen in Table\,\ref{tab:experiments}.

The absolute calibration of molecular fluxes (such as CO) has uncertainties, usually estimated to be around 15\%. For atoms, there are underlying assumptions that enlarge the absolute calibration error. However, the reproducibility of the fluxes have an accuracy of less than 5\%. From one product to another, the main source of uncertainty is the knowledge of ionisation cross-section. Thus, if the relative calibration between compounds may vary \replaced[id=VS]{approximately by}{of about} 15\%, the quantity of the same species is accurately known and can be compared from one experiment to another.

For this project, we performed \added[id=VS]{the} CO hydrogenation experiments shown in Table~\ref{tab:experiments}. Figure~\ref{fig:reactions} represents the visual form of these experiments. The notation is \added[id=VS]{as follows}: if two species are deposited together at the same time (co-deposition), we put both species in curly brackets $\{$CO + H$\}$, and if there was a sequential deposition (one after another), it is noted as $\{$CO$\}$ + $\{$H$\}$. 

\begin{table}
    \caption{List of experimental results from TPD at the end of the deposition.}
    \label{tab:experiments}
    \center
    \begin{adjustbox}{width=1\textwidth}
    \begin{tabular}{*{12}{c}}
        \hline 
       \# & Experiments & Deposit Time & Ts & $\phi_{CO}$ & $\phi_{H}$ & $N_{CO}$ & $N_{H_2CO}$ & $N_{CH_3OH}$ & Radical & COMs & Missing\\
          & & (minutes) & (K) & (ML/s) & (ML/s) & (ML) & (ML) & (ML) & (ML) & & (ML) \\
         \hline
        1&$ $\{CO\} & 10 & 10 & $1.67(-3)$ & $0$ & 1.0 & 0 & 0 & 0 & No & 0 \\
        2&\{CO+H\}\textsuperscript{\emph{a}} & 15 & 10 & $1.67(-3)$ & $6.67(-3)$ & 0.73&0.11 & 0.097 & $<$\replaced[id=BH]{0.1}{0.05}  & No & 0.56 \\
         \hline
         3 & $\{$CO$\}$ & 60 & 15 & $1.85(-3)$ & 0 & 6.667 & 0 & 0 & 0 & No & 0 \\
         4\textsuperscript{\emph{d}} & $\{$CO+H$\}$\textsuperscript{\emph{a}} & 60 & 9 & $1.85(-3)$ & $3.34(-3)$ & 4.063 & 0.169 & 0.039 & $<$\replaced[id=BH]{0.1}{0.05} & No & 2.396 \\
         5\textsuperscript{\emph{d}} & $\{$CO+H$\}$\textsuperscript{\emph{a}} & 60 & 10 & $1.85(-3)$ & $3.34(-3)$ & 4.478 & 0.105 & 0.031 & $<$\replaced[id=BH]{0.1}{0.05} & No & 2.053 \\
         6\textsuperscript{\emph{d}} & $\{$CO+H$\}$\textsuperscript{\emph{a}} & 60 & 11 & $1.85(-3)$ & $3.34(-3)$ & 5.948 & 0.053 & 0.021 & $<$\replaced[id=BH]{0.1}{0.05} & No & 0.645 \\
         7\textsuperscript{\emph{d}} & $\{$CO+H$\}$\textsuperscript{\emph{a}} & 60 & 12 & $1.85(-3)$ & $3.34(-3)$ & 6.448 & 0.036 & 0.015 & $<$\replaced[id=BH]{0.1}{0.05} & No & 0.167 \\
         8 & $\{$CO+H$\}$\textsuperscript{\emph{a}} & 60 & 15 & $1.85(-3)$ & $3.34(-3)$ & 6.593 & 0.018 & 0.010 & $<$\replaced[id=BH]{0.1}{0.05} & No & 0.047 \\
         \hline
         9 & $\{$CO$\}$+$\{$H$\}$\textsuperscript{\emph{b}} & 9+120\textsuperscript{\emph{c}} & 9 & $1.85(-3)$ & $3.34(-3)$ & 0.225 & 0.030 & 0.120 & $<$\replaced[id=BH]{0.1}{0.05} & Yes & 0.625 \\
         10\textsuperscript{\emph{d}} & $\{$CO$\}$+$\{$H$\}$\textsuperscript{\emph{b}} & 9+75\textsuperscript{\emph{c}} & 9 & $1.85(-3)$ & $3.34(-3)$ & 0.321 & 0.0455 & 0.095 & $<$\replaced[id=BH]{0.1}{0.05} & No? & 0.539 \\
         \hline
    \end{tabular}
    \end{adjustbox} 
    \small
    \textsuperscript{}\# and Experiments columns are the index number and the type of the experiment that was done (respectively), ''Deposit time''~- the time frame of the experiment (the time of the deposition of reactants) and Ts is the deposition temperature of reactants. $\phi_{CO}$ and $\phi_{H}$~- the fluxes of CO and H respectively. $N_{CO}$, $N_{H_2CO}$, $N_{CH_3OH}$~- quantities of CO, H$_2$CO and CH$_3$OH (respectively) on the surface in monolayers (ML). ''Radical'' indicates the upper limit of both HCO and CH$_2$OH constrained with RAIRS. ''COMs''~- the presence of complex organic molecules in the TPD traces. ''Missing''~- the quantities of molecules that are desorbing by chemical desorption or the formation of unknown molecules.
    \textsuperscript{\emph{a}} Codeposition of CO and H;
    \textsuperscript{\emph{b}} Deposition of CO followed by deposition of H; \textsuperscript{\emph{c}} deposition time of CO + deposition time of H; \textsuperscript{\emph{d}} Experiments used for comparison in \citealt{husquinet2025neon}.
\end{table}

Here, CO deposition (Experiment \#1) was made for calibration purposes and was taken as a starting point for all experiments below.  Experiment \#2 - CO hydrogenation in sub-monolayer coverage regime that leads to an enhanced chemical desorption. Experiments \#4-8 show intermediate thicknesses of ice on the surface, and the surface temperature in these experiments varied between 9 and 15\,K. Experiments \#9 and \#10 are long hydrogenation of pre-deposited CO ice. This set of experiments has been built to increase the number of experimental constraints (surface densities and temperatures) while keeping the experimental conditions coherent.

The ''Missing'' column in Table \ref{tab:experiments} indicates the amount of CO that is not present compared to a reference experiment. This is based on the assumption that one molecule of CO is required to produce one molecule of \ce{H2CO} and one molecule of CO for one molecule of \ce{CH3OH}. As a result, the missing column reflects chemical desorption \cite{Dulieu2013}, or potential forms of complex organic molecules (COMs), such as methoxymethanol or glycolaldehyde. They are easily identified by their fragments \ce{HCO^+} (m/z=29) or \ce{CH2OH^+} (m/z=31) desorbing at temperatures higher than 150\,K, in addition to higher masses fragmentation patterns (such as m/z = 61 or 62). The determination of the chemical network of such secondary products that are only detectable in very specific conditions (long hydrogenation) is not in the scope of the current study.

\begin{figure}[h!]
    \centering
    \includegraphics[width=0.7\columnwidth]{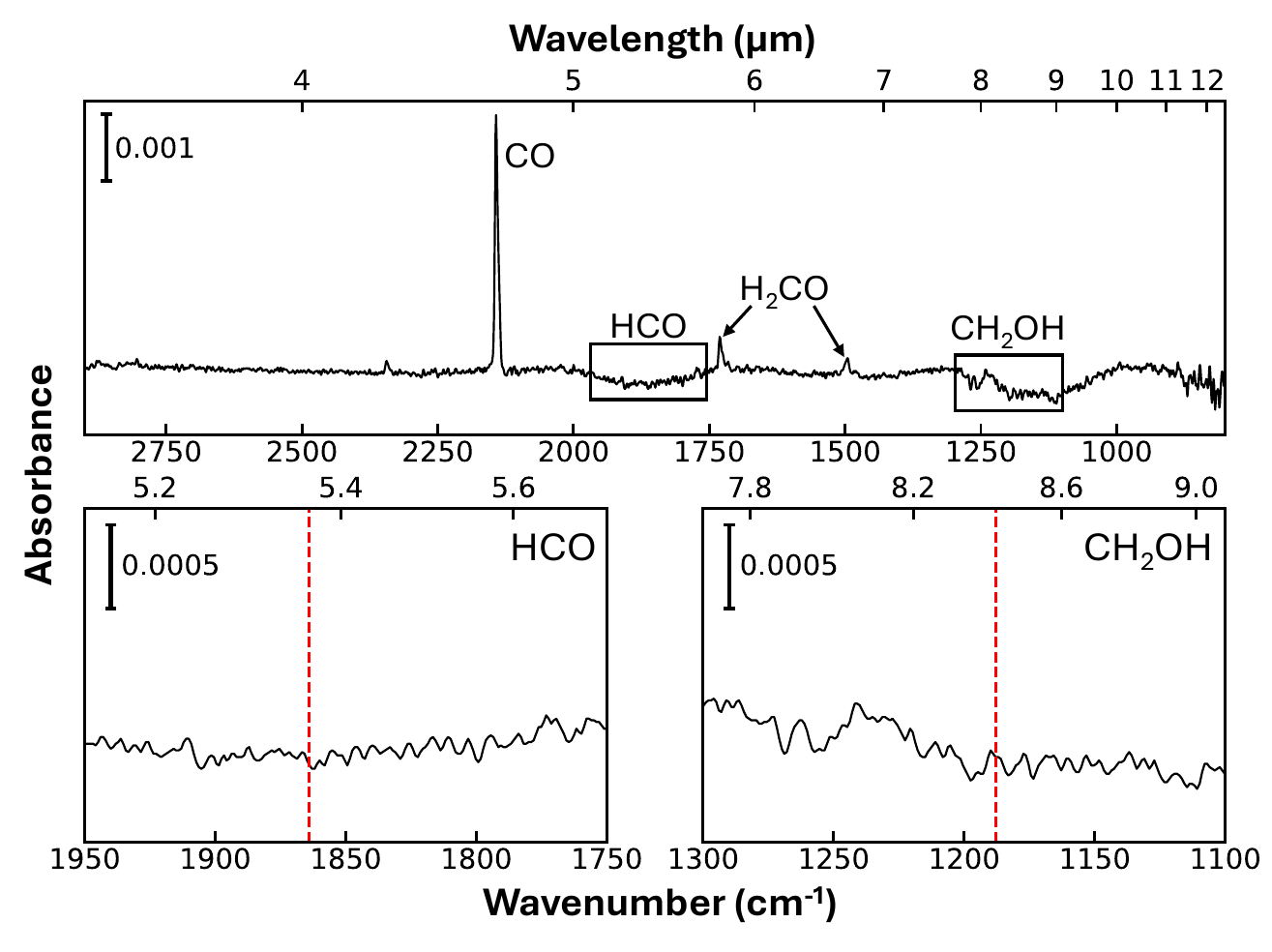}
    \caption{
    The infrared absorption spectra from the {CO+H} co-deposition experiment (experiment \#4 from table\,\ref{tab:experiments}). The top panel shows the global spectra, identifying CO at $2140$\,cm$^{-1}$ and \ce{H2CO} at $1730$\,cm$^{-1}$ and $1494$\,cm$^{-1}$, while \ce{CH3OH} remains undetected due to its low abundance. The lower panel focuses on \ce{HCO} radical at $1864$\,cm$^{-1}$ and \ce{CH2OH} at $1188$\,cm$^{-1}$ \cite{butscher2015formation}, which are not detected in our experiments.}
    \label{fig_exp: IR spectra and radicals}
\end{figure}

The ''Radical'' column refers to the presence of \ce{HCO} and \ce{CH2OH} in our experiments. 
\replaced[id=BH]{A conservative value of 0.1~ML was adopted as the upper limit, based on the detection limit for the \ce{CO} $\nu_1 = 2139$~cm$^{-1}$ band. As illustrated in Figure~\ref{fig:COMLSpec}, the IR spectra obtained during the \ce{CO} deposition experiment reveal the emergence of the \ce{CO} band starting at scan~2, which corresponds to approximately 0.2~ML of \ce{CO}. Given the CO band strength $A_{\nu_1}^{\ce{CO}}=8.8\times10^{-18}$~cm/molec for an IR resolution of 2~cm$^{-1}$ and a surface temperature of 10~K \cite{2023MNRAS.522.3145G}, the upper limit for \ce{HCO} detection is estimated to be $\sim0.1$~ML, assuming a band strength of $A_{\nu_3}^{\ce{HCO}}=1.8\times10^{-17}$cm/molec\cite{2017AcSpA.187...39R}. For the case of \ce{CH2OH}, since no band strength was available in the literature, the upper limit value as \ce{HCO} was adopted.}{A conservative value of 0.05\,ML was adopted as the upper limit because we did not detect any radicals in any of the experiments presented in Table\,\ref{tab:experiments}.} 
Figure~\ref{fig_exp: IR spectra and radicals} shows, as an example, the regions of the IR spectrum where these radicals should be detected. 
\replaced[id=BH]{As observed in the lower panel of Figure~\ref{fig_exp: IR spectra and radicals}, neither the \ce{HCO} nor the \ce{CH2OH} bands are visible. This observation is consistent across other spectra and experiments. These radicals are not part of the ''missing'' molecules because we assumed their reactions with H atoms (or other radicals) occurs too rapidly to be detected by our experimental setup. A few experimental studies have detected \ce{HCO} or \ce{CH2OH} in thick ice ($>100$~ML) by isolating them in chemically inert molecular matrices\cite{2002A&A...386.1129Z,butscher2015formation}. Under our experimental conditions, we over-hydrogenate \ce{CO} with a flux ratio of $\phi_{\ce{CO}}/\phi_{\ce{H}} = 0.25-0.55$, which gives the H atoms time to scan and react with all radicals present on the surface.}{These radicals are not part of the ''missing'' molecules, but their reactivity with H atoms (or other radicals) occurs too rapidly to be detected with our experimental setup.}
Finally, using the combination of TPD and FT-RAIRS, we can constrain the evolution of the different molecules involved in the {\ce{CO + H}} reactive systems. The quantity of products observed (or not) in addition to the condition used are summarized in Table\,\ref{tab:experiments}. 

\section{Rate-equations approach}\label{sec:re}

There are different theoretical methods for studying the chemical evolution of ice and molecules on the surface of ice, such as quantum calculations, Kinetic Monte Carlo calculations and classical rate equation approaches~\cite{2024arXiv240706657C}. In the scope of this project we decided to base our theoretical studies on the Rate Equation (RE) approach and use astrochemical codes that utilize this approach: MONACO~\cite{2013ApJ...769...34V}, NAUTILUS~\cite{2016MNRAS.459.3756R} and pyRate~\cite{sipila2015benchmarking}.

The RE approach is the standard method for modelling complex gas-dust chemistry (e.g., ~\citealt{hasegawa1992models,1993MNRAS.263..589H, 2008ApJ..682..283G, 2017ApJ...842...33V, 2016MNRAS.459.3756R}). This type of code takes into account the rates and probabilities of accretion of atoms and molecules on dust particle surfaces, their migration on the dust surface, the probabilities of molecule formation, and the probabilities of the molecule desorption to the gas phase. All codes used in this paper can calculate the gas-phase chemistry and chemistry on the dust surfaces with different conditions of the environment and different hypotheses.

The kinetic equations for the gas-phase species that are used in this method have the general form:

\begin{equation}\label{kinematic}
\frac{dn_{i,gas}}{dt}=\sum\limits_{j,k}k_{j,k}n_{j}n_{k} + \sum\limits_{l}k_{l}n_{l} - n_{i}\left[\sum\limits_{m}k_{im}n_{m} + \sum\limits_{n}k_{n}\right] + k_{des}n_{i,s} - k_{acc}n_{i},
\end{equation}
where $n_{i}$~--- abundance of i-th compound that we need to know; $k_{j,k}$ and $k_{l}$, $k_{im}$ and $k_{n}$, $k_{des}$, $k_{acc}$~---  the rate coefficients of the reactions of formation (first pair), molecule destruction (second pair), desorption, and accretion, respectively. For the equations of chemical kinetics on the dust surface, the signs (+ or -) of the $k_{des}n_{i,s}$ and $k_{acc}n_{i}$ are reversed:
\begin{equation}\label{kinematic}
\frac{dn_{i}}{dt}=\sum\limits_{j,k}k_{j,k}n_{j}n_{k} + \sum\limits_{l}k_{l}n_{l} - n_{i}\left[\sum\limits_{m}k_{im}n_{m} + \sum\limits_{n}k_{n}\right] - k_{des}n_{i,s} + k_{acc}n_{i}.
\end{equation}

Despite the advantages of this approach (ease of use, relatively fast calculations), there are also disadvantages that can be ignored in the present study. For example, the RE method is a macroscopic approach that does not take into account all microscopic details on the dust surface and can not track the position of atoms and molecules on the surface of dust. It can lead to the overproduction of some molecules, as soon as method operates with the averaged values of the species (it is especially important when the number of species on the dust surfaces is low)~\cite{1994ApJ...421..206C}. For now, our main goal is to try to show that data from astrochemical codes based on the RE approach can be used for comparison with laboratory results: they can reproduce results that we see with simple experiments (with a given level of uncertainties) and later can be used for prediction of products that can be possibly found in experiments. 

Below we describe what was adapted in the codes to mimic laboratory experiments. In principle, we did not change the codes in technical terms (functions, used packages, additional parts of the code), or physical terms (such as adding or removing a process), but changed only the input parameters. For all codes general rules were:
\begin{enumerate}
    \item Gas phase interactions are neglected: all gas phase reactions are turned off, except for the freeze-out of species from gas phase to the surface of dust;
    \item Only thermal and chemical desorption are allowed in all codes;
    \item In all codes we use the same parameters for grains: each grain is assumed to be a uniform spherical particle with a radius of 0.1\,$\mu$m, and its number density is calculated using the gas to dust ratio of 1\%;
    \item The surface of dust particles is assumed to be compact with density of 1.5$\times$10$^{15}$\,sites\,cm$^{-2}$ (or about 10$^6$\,sites per grain). Grain material density is set to $3$\,g/cm$^3$. These are the standard parameters for all the codes, and they were not changed in the present study~\cite{2009ApJ...691.1459V,2017ApJ...842...33V,2016MNRAS.459.3756R,sipila2015benchmarking};
    \item Gas-phase chemistry and chemistry on the dust surfaces are connected via accretion and desorption processes, sticking probability is set to 1 in the case of MONACO (for most molecules, except for H and H$_2$, for them sticking coefficients are calculated separately) and NAUTILUS codes. In the case of pyRate, the sticking coefficient of \ce{H} is computed following the Cuppen equations\cite{cuppen2010h2}, with probabilities of $0.42$ at a gas temperature of $293$\,K and a dust temperature of $10$\,K. For \ce{H2} and \ce{CO}, the sticking coefficients are derived using the He equations\cite{he2016sticking}, yielding sticking probabilities of $0.62$ and $1.0$, respectively. All other molecules are assumed to have a sticking probability of 1. For all codes, the species accreted from the gas phase can react through the Eley-Rideal mechanism with species on the surface, or diffuse via thermal hopping. In case of \ce{H} and \ce{H2}, diffusion by thermal hopping or tunnelling was considered.
    \item Binding-to-diffusion energy ratios are different from one code to another. For MONACO code E$_{d}$ / E$_{b}$ = 0.5 for the surface (and 1.0 for the bulk), for NAUTILUS~- 0.4 (and 0.8 for the bulk) and for pyRate~- 0.55 (with no diffusion for the bulk).
\end{enumerate}

\section{MONACO code and results}\label{sec:monaco}

MONACO code is a gas-grain chemical model based on the RE approach for treating gas-phase reactions~\cite{2013ApJ...769...34V} and modified RE approach for reactions inside the ice bulk~\cite{2008A&A...491..239G, 2009ApJ...700L..43G}. This code implements a multi-phase approach for studying the chemical evolution of species: gas-phase chemistry, chemistry on the surface and chemistry inside the interstellar ice. Kinetic equations for chemistry on the grain surfaces take into account the complex structure of ice mantles, which consists of the bulk phase and ice surface, migration of species inside the ice, and much more. The detailed description of the 3-phase treatment of chemistry can be found in~\citealt{2017ApJ...842...33V}.

The MONACO code operates with relative abundances of species. To calculate the number of molecular monolayers on the dust surface we need to recalculate these abundances using the assumption:

\begin{equation}\label{monolayer} 
\centering
N_{ML} = X \cdot \frac{n}{n_{dust}\cdot N_{sites}},
\end{equation} 
where $X$ is the relative abundance of the molecule $n(X)/n_H$ (here $n_H$ is the total abundance of hydrogen equal $n(H) + 2n(H_2)$) , $n$ - number density of gas in the system, $n_{dust}$~ - number density of dust grains in the system, $N_{sites}$~ - number of sites on the surface. For the proper calculation of $n(X)$, we need to take into account species in the bulk and on the ice surface as a sum. Here and after we will denote all the molecules as follows: X (if species do not have a prefix) as a molecule in the gas phase, gX~--- molecule on the grain surface, mX~--- molecules from bulk and surface together (gX + bX). 

In the MONACO code we use the KIDA database~\cite{2012ApJS..199...21W, 2015ApJS..217...20W}, which takes into account not only gas-phase reactions but also reactions on the surface of dust particles and their ice mantles. The main reactions leading to the formation of methanol (and important in $\{$CO+H$\}$ system) are presented in Table~\ref{tab:monaco_ch3oh}.

\subsection{Experiment \#1 and \#2: CO deposition on the surface and {CO+H} co-deposition with reference fluxes}\label{subsec:exp_1_2}

We started our calculation tests with simple CO deposition on the grain surface following the protocol of experiments described before. The main physical parameters that we need to know and put into the system are the gas and dust temperatures (\replaced[id=VS]{$T_{gas}$}{T$_{gas}$} and \replaced[id=VS]{$T_{d}$}{T$_{d}$}, respectively), the gas abundance (\replaced[id=VS]{$n(H)$}{n(H)}) and the initial abundance of CO. All other parameters will remain untouched. 

The CO gas-phase abundance is derived using the standard formula of accretion, assuming that the flux that we see in the laboratory is indeed our accretion rate:

\begin{equation}\label{accrate}
 k_{acc}=n\,\sigma_{d}\,\upsilon S(T_{gas},T_{d}),
\end{equation}
where $n$~--- the number density of particles in the gas in cm$^{-3}$; $\sigma_{d}$~--- the effective cross-section of the dust grains, close to its geometric area in cm$^{2}$; $\upsilon$~--- the average thermal velocity of the gas particles in cm/s; $S$~--- the sticking coefficient, dimensionless. In the case of CO deposition for our tests, we have: \replaced[id=VS]{$S(T,T_{d}$)}{S(T,T$_{d}$}) equals to unity (all particles stick to the surface), \replaced[id=VS]{$T_{gas}$}{T$_{gas}$} = 300~K, \replaced[id=VS]{$T_{d}$}{T$_{d}$} = 10~K, and \replaced[id=VS]{$k_{acc (CO)}$}{k$_{acc (CO)}$} = 1.67$\times10^{12}~\text{molecules cm}^{-2}\text{s}^{-1}$, from our experimental conditions. Hence, the number density of CO in the gas phase can be calculated as follows:
\begin{equation}
n(CO) = \frac{1.67\times10^{12} \text{molecules cm}^{-2}\text{s}^{-1}}{\pi \cdot a_{gr}^{2}\cdot \upsilon} = 1.15 \times 10^{8}~\text{cm}^{-3}.
\end{equation}

All of that leads us to the first model with given parameters:
\begin{enumerate}
    \item $n = 1.15 \times 10^{8}~\text{cm}^{-3}$ - as gas number density ($n_H$);
    \item \replaced[id=VS]{$T_{gas}$}{T$_{gas}$} = 300~K, \replaced[id=VS]{$T_{d}$}{T$_{d}$} = 10~K;
    \item initial gas-phase abundance: \replaced[id=VS]{$n(CO)$}{n(CO)}/$n_H$ = 1.
\end{enumerate}

\begin{figure}[h!]
        \includegraphics[width=0.8\linewidth]{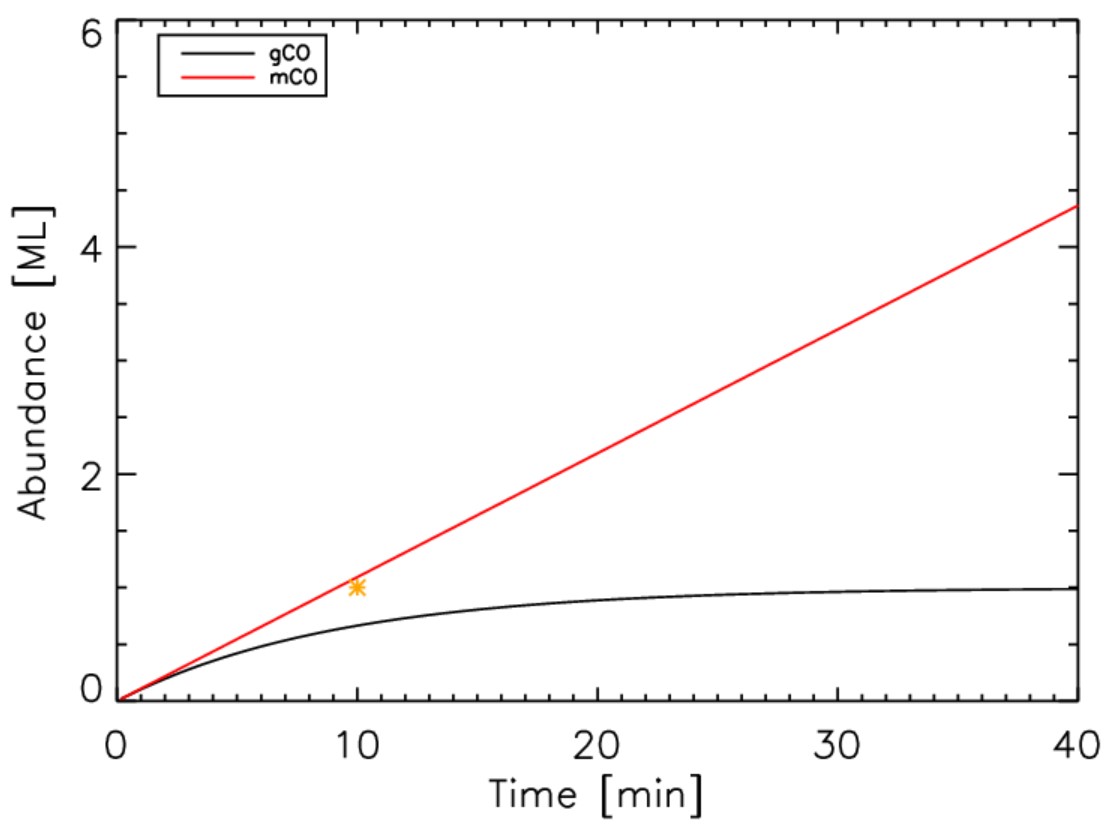}
    \caption{Experiment \#1: Growth of the CO ice on the surface, MONACO code. mCO denotes molecules from bulk and surface together (red line). Saturation of the chemical layer (black line), where the active layer is set to be equal to 1. The orange star represents results from the experiments, where we obtain 1~ML of CO on the surface in 10 minutes.}
    \label{fig:monaco_co_only}
\end{figure}

Figure~\ref{fig:monaco_co_only} shows results for modelling CO deposition on the surface of a dust grain: growth of the CO ice and saturation of the chemical layer which is equal to 1~ML. These results fit great with experimental data and give us 1.1~ML of CO ice in 10 minutes.

After that, we added the hydrogen flux to our system, taking into account values that were given in the Table~\ref{tab:experiments} for Experiment \#2. Hence, the number density of H in the system will be:
\begin{equation}
n(H) = \frac{6.67\times10^{12} \text{molecules cm}^{-2}\text{s}^{-1}}{\pi \cdot a_{gr}^{2}\cdot \upsilon \cdot 0.8} = 1.07 \times 10^{8}~\text{cm}^{-3},
\end{equation}
where 0.8 is the sticking coefficient of H (calculated in the code~\cite{2017ApJ...842...33V}, using a formula from~\citealt{1979ApJS...41..555H}).

All of that gives us the following set of parameters:
\begin{enumerate}
    \item $n = 1.074 \times 10^{8}~\text{cm}^{-3}$ - as gas number density ($n_H$);
    \item \replaced[id=VS]{$T_{gas}$}{T$_{gas}$} = 300~K, \replaced[id=VS]{$T_{d}$}{T$_{d}$} = 10~K;
    \item initial gas-phase abundances: \replaced[id=VS]{$n(H)$}{n(H)}/$n_H$ = 1, \replaced[id=VS]{$n(CO)$}{n(CO)}/$n_H$ = 1.07. It is important to note that one must carefully recalculate this parameter every time the density changes;
    \item models were calculated in two regimes: tunnelling of light species (H) is turned on and off;
    \item methanol-production reaction chain, that presented in Table~\ref{tab:monaco_ch3oh}, is used for calculating the chemistry.
\end{enumerate}

\begin{table*}
\caption{\label{tab:monaco_ch3oh}Table of reactions included in the methanol-formation chain. Here, $E$ is the energy of the process (binding energy or activation energy depending on the type of the process) in each of the codes: $E_{MC}$, $E_{NT}$, $E_{PR}$ for energies in MONACO, NAUTILUS and pyRate codes, respectively. x denotes reactions that are not included in the code's chemical network.}
\centering
\begin{tabular}{lccc}
\hline\hline
Reaction& $E_{MC}$ & $E_{NT}$ &$E_{PR}$\\
\hline
CO $\rightarrow$ gCO \textsuperscript{\emph{a}} 	&	 0.0  	& 0.0	& 0.0	\\
H $\rightarrow$ gH \textsuperscript{\emph{a}}	&	 0.0  	&	0.0	& 0.0	\\
H$_2$ $\rightarrow$ gH$_2$ \textsuperscript{\emph{a}}	&	 0.0  	&	0.0	& 0.0		\\ \hline
gCO $\rightarrow$  CO \textsuperscript{\emph{b}}	&	 1322.0  	&	1300.0	& 1150.0	\\
gHCO $\rightarrow$ HCO \textsuperscript{\emph{b}}	&	 1600.0  	&	2400.0	& 1600.0	\\
gH$_2$CO $\rightarrow$   H$_2$CO \textsuperscript{\emph{b}}	&	 4009.0  	&	4500.0	& 2050.0	\\
gCH$_3$O  $\rightarrow$   CH$_3$O \textsuperscript{\emph{b}}	&	 4400.0  	&	4400.0	& 3800.0	\\
gCH$_2$OH $\rightarrow$   CH$_2$OH \textsuperscript{\emph{b}}	&	  5080.0  	&	4400.0	& 5084.0	\\
gCH$_3$OH $\rightarrow$   CH$_3$OH \textsuperscript{\emph{b}}	&	 5740.0  	&	5000.0	& 5534.0		\\
gH $\rightarrow$   H \textsuperscript{\emph{b}}	&	 450.0  	&	650.0	& 450.0	\\
gH$_2$ $\rightarrow$   H$_2$ \textsuperscript{\emph{b}} 	&	  430.0  	&	440.0	& 500.0	\\ \hline
gH + gH $\rightarrow$ gH$_2$ 	&	 0.0  	&	0.0	&	0.0	\\
gH + gCO $\rightarrow$ gHCO 	&	  2320.0  	&	2500.0	&	1730.0	\\
gH + gHCO $\rightarrow$ gH$_2$CO 	&	 0.0 	&	0.0	&	0.0	\\
gH + gHCO $\rightarrow$ gCO + gH$_2$ 	&	 0.0  	&	x	&	0.0	\\
gH + gH$_2$CO $\rightarrow$ gHCO + gH$_2$ 	&	  2960.0  	&	1740.0	&	3000.0	\\
gH + gH$_2$CO $\rightarrow$ gCH$_2$OH 	&	 4500.0  	&	5400.0	&	5160.0	\\
gH + gH$_2$CO $\rightarrow$ gCH$_3$O 	&	 2320.0  	&	2200.0	&	2000.0	\\
gH + gCH$_2$OH $\rightarrow$ gCH$_3$OH 	&	 0.0  	&	0.0	&	0.0	\\
gH + gCH$_3$O $\rightarrow$ gCH$_3$OH 	&	 0.0  	&	0.0	&	0.0	\\
gH + gCH$_2$OH $\rightarrow$ gH$_2$CO + gH$_2$ 	&	 0.0  	&	x	&	0.0	\\
gH + gCH$_3$O $\rightarrow$ gH$_2$CO + gH$_2$ 	&	 0.0  	&	x	&	0.0	\\
gH + gCH$_3$OH $\rightarrow$ gCH$_2$OH + gH$_2$ 	&	 4380.0  	&	2600.0	&	3620.0	\\
gH + gCH$_3$OH $\rightarrow$ gCH$_3$O + gH$_2$ 	&	 6640.0  	&	4000.0	&	5560.0	\\ \hline

\end{tabular}

\textsuperscript{\emph{a}}Freeze-out on the surface;\textsuperscript{\emph{b}}Desorption from the surface
\end{table*}

Table~\ref{tab:monaco_exp2} displays the result of modelling for Experiment \#2 with two regimes: with tunnelling of light species (H) on and off. As can be seen from the results, if tunnelling diffusion for light species is turned on, all hydrogen atoms quickly transform into molecular hydrogen and almost no $\{$CO+H$\}$ chemistry occurs on the surface of dust particles. According to~\citealt{2017MolAs...6...59S}, tunnelling of H atoms plays a significant role in the chemistry of H$_2$ on the surface of ASW with temperatures around 10~K and lower. In addition to that, they mention that all their calculation results for binding energies and rate constants are strongly dependent on the morphology and structure of the surface. In our models, the dust grain and the ice surface are uniform, so we suggested that \replaced[id=VS]{$T_{dust}$}{T$_{dust}$}=10~K could be too high for the tunnelling to dominate over thermal hopping. Indeed, the so-called crossover temperature $T_c$ at which the tunnelling effect is similar to the thermal crossing has been calculated to be lower than 10~K in the case of H diffusion on amorphous solid water ice~\cite{Nyman2021}. As one can see, for our experiments the thermal hopping of hydrogen seems to be a better description than tunnelling, even though the substrate is not amorphous solid water ice but a gold surface (not monocrystalline). 

\begin{table*}[h!]
\caption{Modelling results from MONACO code for Experiment \#2 with two regimes: tunnelling of light species on and off. Monolayers are calculated for the sum of bulk and surface quantities of molecules.}             
\label{tab:monaco_exp2}    
\centering  
\begin{tabular}{l c c c c c c}  
\hline                 
 & CO & HCO & H$_2$CO & CH$_2$OH & CH$_3$O & CH$_3$OH \\ 
& \multicolumn{6}{c}{(ML)}\\ 
\hline  
Experimental results & 0.730 & - & 0.11 & - & - & 0.097\\
(from Table~\ref{tab:experiments}) &&&&&& \\ \hline
Modelling results  &&&&&& \\ \hline
Tunneling diffusion of & 1.664  & 3.217(-9) & 2.141(-3) & 3.885(-15) & 4.590(-12) & 2.740(-6) \\
light species on & & & & & & \\
Tunneling diffusion of & 0.982 & 1.400(-3) & 0.298 & 1.224(-4) & 4.098(-4) & 0.139 \\
light species off &&&&&& \\ 
\hline                          
\end{tabular}
\end{table*}

\begin{figure}
     \includegraphics[width=0.8\linewidth]{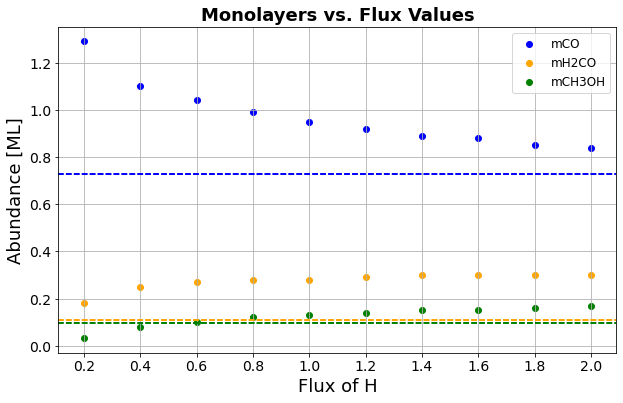}
    \caption{H-atom flux variation vs number of monolayers for CO, H$_2$CO and CH$_3$OH. Dashed lines represent the laboratory results for each molecule, dots note the amount of monolayers for each molecule. The base H-atom flux was taken as $n = 1.074 \times 10^{8}~\text{cm}^{-3}$ (corresponds to 1.0 on the X axis of the graph)}
    \label{fig:monaco_flux_var}
\end{figure}

One of the possible solutions for controlling the molecule production here is the variation of the H-atom flux in the system. Figure~\ref{fig:monaco_flux_var} shows how the H-flux variation in the case of Experiment \#2 influences the chemical composition of dust ices. Calculations show that the higher we take the flux - the stronger the destruction of CO, and in most cases we see the overproduction of H$_2$CO and CH$_3$OH in comparison with experiments. 

In general, modelling results for Experiments \#1 and \#2 are close to the ones that we see in the \replaced[id=VS]{laboratory}{lab}, even though the production effectiveness of all molecules appears to be a little higher than in the experiments. Indeed, the missing part is underestimated, indicating that the chemical desorption is possibly also undervalued.

\subsection{Experiments \#4-\#8: {CO} deposition and {CO+H} co-deposition}\label{subsec:exp_3_8}

We do not show the results from modelling of the Experiment \#3 here because in principle they are the same as the results for Experiment \#2: the growth of mCO follows the same linear trend and gives us 7.2~ML of CO in 60 minutes. In the set of experiments (\#4-\#8) we studied the co-deposition \{CO+H\} with different temperatures of the surface (from 9 to 15~K).

Using the same Equation~\ref{accrate} we derived new abundances for H atoms and CO molecules in the system, using new laboratory fluxes from Table~\ref{tab:experiments}: $n(H) = 5.3 \times 10^{7}~\text{cm}^{-3}$, $n(CO) = 1.26 \times 10^{8}~\text{cm}^{-3}$.

\begin{figure}
     \includegraphics[width=0.7\linewidth]{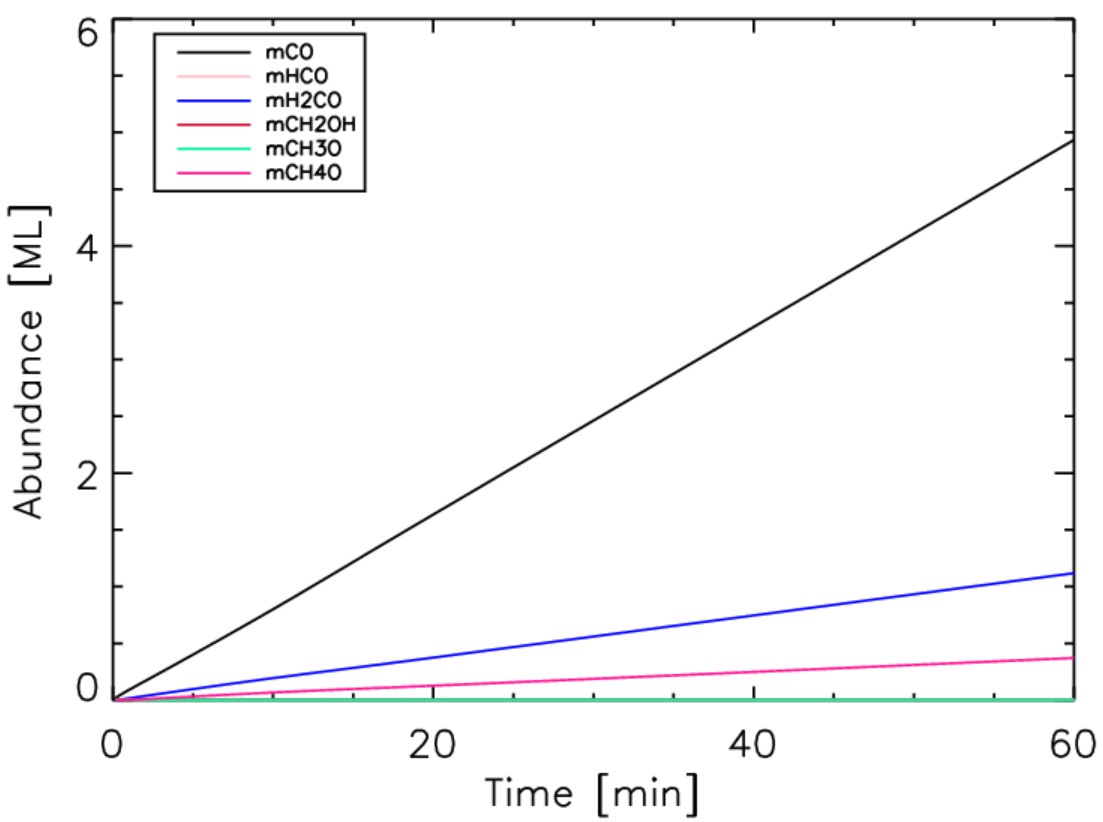}
    \caption{Modelling results for Experiment \#5 (co-deposition of CO and H - \{CO+H\} - at surface temperature of 10~K), the MONACO code. mX (where X is the chemical formula of the compound) denotes molecules from bulk and surface together.}
    \label{fig:monaco_exp5}
\end{figure}

Here, we started our calculating with Experiment~\#5 hence deposition at 10~K is considered to be standard for the experiments. Figure~\ref{fig:monaco_exp5} shows results for modelling co-deposition of CO and H for the system with the following parameters:
\begin{enumerate}
    \item $n = 5.3 \times 10^{7}~\text{cm}^{-3}$ - as gas number density ($n_H$);
    \item \replaced[id=VS]{$T_{gas}$}{T$_{gas}$} = 300~K, \replaced[id=VS]{$T_{d}$}{T$_{d}$} = 10~K;
    \item initial gas-phase abundances: \replaced[id=VS]{$n(H)$}{n(H)}/$n_H$ = 1, \replaced[id=VS]{$n(CO)$}{n(CO)}/$n_H$ = 2.38;
    \item methanol-production reaction chain, that presented in Table~\ref{tab:monaco_ch3oh}.
\end{enumerate}

From modelling results we see the overproduction of CO, H$_2$CO, and CH$_3$OH molecules. In our study we use the same treatment of reactive desorption, noted RD, that is used in ~\citealt{2017ApJ...842...33V}. In this work they implemented the results from experiments of~\citealt{2016A&A...585A..24M} and described the dependence of the RD efficiency ($R_{\text{RD}}$) on important surface parameters (the composition of the surface, surface reaction exothermicity, and binding energies of reaction products):
\begin{equation} \label{eq: chemical desorption}
    R_{\text{RD}} = \exp\left(-\epsilon \frac{E_b \times DF}{\Delta H}\right),
\end{equation}
where \replaced[id=VS]{$E_b$}{E$_b$} is the binding energy, $DF$ is the number of vibrational modes in the molecule–surface bond system, \added[id=VS]{and }$\epsilon$ is the fraction of kinetic energy kept by the reaction product.

In their study, \citealt{2017ApJ...842...33V} show the same problem with overproduction of H$_2$CO and CH$_3$OH on the surface of dust particles in comparison with observations of interstellar ices, referring to the semi-empirical nature of expression for RD that was based on the limited set of experiments (in addition, the same issue was discussed in~\citealt{2025MNRAS.537.3686P}). Moreover, some parameters that this expression uses are arguable: for example, the binding energies of certain molecules are still unknown, which also requires the revision of the parameters of the chemical network used in the code and leads to the extrapolation of results from well-known systems to poorly known ones. Thus, one of the possible explanations for the overproduction of H$_2$CO and CH$_3$OH that we see in our results is the insufficiently accurate treatment of RD for these species and their reaction systems. 

One way to get more precise results would be the changing of the binding energy of H atoms to control the hydrogenation rate of all reactants and activation energy of the reaction CO + H $\rightarrow$ HCO, the key step in the H$_2$CO production reaction chain. The activation energy (\replaced[id=VS]{$E_a$}{E$_a$}) of reactions is one of the crucial issues in laboratory and theoretical astrochemistry. CO + H $\rightarrow$ HCO reaction was well studied by several groups, and \replaced[id=VS]{$E_a$}{E$_a$} values for this reaction vary from 1635~K~\cite{2019A&A...627A...1Q} to 2320~K~\cite{2013ApJ...765...60G}. The same uncertainties come for the binding energy of hydrogen: in the chemical network that we used in past models and studies it is assumed that \replaced[id=VS]{$E_b$}{E$_b$}(H) = 450~K (from the  OSU gas-grain code of Eric Herbst group in 2006, where it was calculated in~\citealt{hasegawa1992models}), but in recent experiments the value 661~K was introduced (~\citealt{2017MolAs...6...59S}). This parameter directly affects the key chemical parameter, the diffusion energy E$_{d}$ (see the discussion in \citealt{Ligterink2025}) via the E$_d$ = $\beta E_b$ relation. In the case of MONACO code, $\beta$ = 0.5.

In our study, we decided to vary both values (\replaced[id=VS]{$E_b$}{E$_b$}(H) and \replaced[id=VS]{$E_a$}{E$_a$}(CO+H)) and see how it affects the modelling results: for hydrogen diffusion estimation, we took values in the range 350~K $<$ \replaced[id=VS]{$E_b$}{E$_b$}(H) $<$ 650~K, for the CO+H reaction, we considered 1700~K $<$ \replaced[id=VS]{$E_a$}{E$_a$}(CO+H) $<$ 2900~K. The first parameter probes the H diffusion while the second affects the H reactivity with CO.

\begin{figure}[h!]
   \begin{minipage}{0.7\linewidth}
     \centering
     \includegraphics[width=\linewidth]{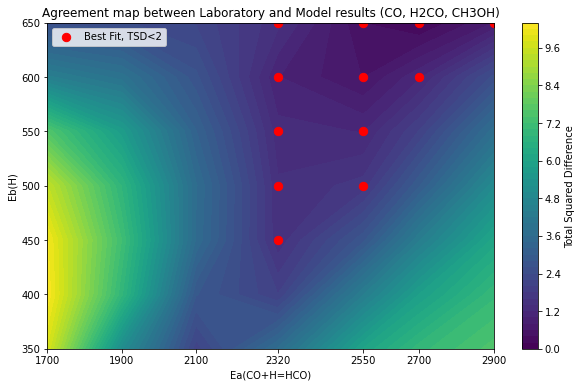}
   \end{minipage}\\
   \begin{minipage}{0.7\linewidth}
     \centering
     \includegraphics[width=\linewidth]{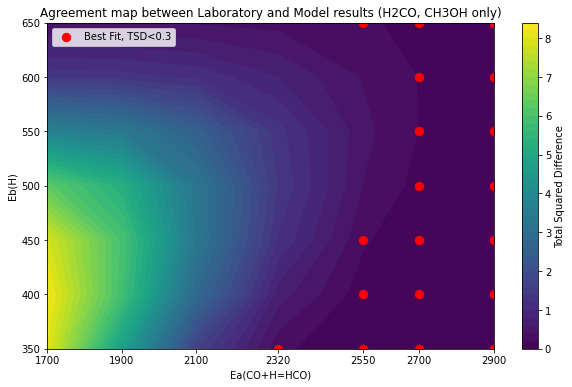}
   \end{minipage}
    \caption{Agreement map between laboratory and modelling results in the case of the MONACO code, Experiment \#5 (co-deposition \{CO+H\} with $T_{surf} = $10~K). Red dots denote the best-fit pair of parameters, where in the upper panel, \replaced[id=VS]{RSS}{TSD} is calculated for all 3 molecular abundances and is lower than 2.0; in the bottom panel, \replaced[id=VS]{RSS}{TSD} is calculated only for H$_2$CO and CH$_3$OH and is lower than 0.3}
    \label{fig:monaco_bestfit}
\end{figure}

Figure~\ref{fig:monaco_bestfit} shows the agreement maps for the calculation grid, where we varied \replaced[id=VS]{$E_b$}{E$_b$}(H) and \replaced[id=VS]{$E_a$}{E$_a$}(CO+H). We calculated the agreement between laboratory and modelling results using the \replaced[id=VS]{Residual sum of squares (RSS) approach}{Total Squared Difference (TSD) approach}. In this approach, we calculated the total squared difference for each combination of parameters:
\begin{equation}
\text{RSS} = \sum^{n}_{i=1}(\text{Model Abundance}_{i} - \text{Laboratory Abundance}_{i})^2,
\end{equation}
where $n$ is the number of molecules considered for the comparison\comment[id=VS]{VS: changed TSD for RSS}. Lower values of this sum indicate that the modelling results better fit with the laboratory measurements, while higher values indicate that the model results deviate significantly from laboratory data. 

Based on the \replaced[id=VS]{RSS}{TSD} approach, we derived the best-fit points, i.e. the set of parameters that show the best agreement between results. The top panel in Figure~\ref{fig:monaco_bestfit} shows the agreement map and best-fit points if \replaced[id=VS]{RSS}{TSD} is calculated for all 3 molecules. The value of CO gives the largest contribution to the \replaced[id=VS]{RSS}{TSD} values, thus we decided to build the additional map only for H$_2$CO and CH$_3$OH (bottom panel in Figure~\ref{fig:monaco_bestfit}). 

Analysis of both best-fit results shows that there is no set of parameters which provides good agreement for all three molecules at once: either we have a high value for CO, or the overproduction of H$_2$CO and CH$_3$OH. Another constraint set by results from the table~\ref{tab:experiments} is that there are no signs of HCO or CH$_3$O in Experiment \#5 both on TPD or IR spectra. Based on this, \replaced[id=VS]{$E_b$}{E$_b$}(H) $\geq$ 550~K are not considered as with these values, we measure large amounts of HCO or CH$_3$O - around 0.4 and 0.2 ML respectively, which is not matching the laboratory results. On the other hand, \replaced[id=VS]{$E_b$}{E$_b$}(H) $<$ 400~K appears to be very low and can lead to high overproduction of CO with larger \replaced[id=VS]{$E_a$}{E$_a$}(CO+H) (values closer to 2900~K)  and high overproduction of H$_2$CO with lower \replaced[id=VS]{$E_a$}{E$_a$}(CO+H) ((values closer to 1700~K). In addition to that, from the agreement maps we note that there is an intersection between good fitting areas when 2320~K $\leq$ \replaced[id=VS]{$E_a$}{E$_a$}(CO+H) $\leq$ 2550~K. Therefore, we arrive at the following ranges: 2320~K $\leq$ \replaced[id=VS]{$E_a$}{E$_a$}(CO+H) $\leq$ 2550~K and 400~K $\leq$ \replaced[id=VS]{$E_b$}{E$_b$}(H) $\leq$ 500~K.

From these values, we choose the point with the lowest \replaced[id=VS]{RSS}{TSD}: \replaced[id=VS]{$E_b$}{E$_b$}(H) = 500~K and \replaced[id=VS]{$E_a$}{E$_a$}(CO+H) = 2320~K, where \replaced[id=VS]{RSS}{TSD} = 1.43 (see Table~\ref{tab:monaco_exp5_5002320}). These energy values are consistent with values obtained by other groups. For example, ~\citealt{2025ApJ...990..163B} calculated \replaced[id=VS]{$E_b$}{E$_b$}(H) = 480~K based on the work of~\citealt{2022ESC.....6..597M}, and \replaced[id=VS]{$E_a$}{E$_a$}(CO+H) = 2320~K. For further analysis, we keep these two values fixed (\replaced[id=VS]{$E_b$}{E$_b$}(H) = 500~K and \replaced[id=VS]{$E_a$}{E$_a$}(CO+H) = 2320~K). 

\begin{table*}[h!]
\caption{Modelling results from the MONACO code for Experiment \#5 (co-deposition \{CO+H\} with $T_{surf} = $10~K): chemical composition of the ice before modifications in the initial chemical network, at the ''best fit point'' (\replaced[id=VS]{$E_b$}{E$_b$}(H) = 500~K and \replaced[id=VS]{$E_a$}{E$_a$}(CO+H) = 2320~K), and after modifications in the chemical network at the ''best fit'' point. Monolayers are calculated summing bulk and surface molecules.}             
\label{tab:monaco_exp5_5002320}    
\centering  
\begin{tabular}{l c c c c c c}  
\hline\hline                 
 & CO & HCO & H$_2$CO & CH$_2$OH & CH$_3$O & CH$_3$OH \\ 
& \multicolumn{6}{c}{(ML)}\\ 
\hline     
Experimental results & 4.478 & - & 0.105 & - & - & 0.031 \\ \hline
Modelling results &&&&&& \\ \hline
Before modifications & 5.005 & 6.708(-3) & 1.134 & 3.323(-4) & 1.521(-3) & 0.376  \\      
''Best fit'' point & 4.699 & 4.092(-2) & 1.202 & 2.354(-3) & 9.864(-3) & 0.454  \\
''Best fit'' point with modifications & 5.545 & 3.900(-2) & 0.306 & 1.454(-3) & 2.035(-3) & 3.242(-2) \\
\hline                                  
\end{tabular}
\end{table*}

In the next step, in an attempt to improve the agreement between experimental and model results, the following parameters have been changed (the initial parameters for these reactions are shown in Table~\ref{tab:monaco_ch3oh}):
\begin{enumerate}
    \item the branching ratios for reactions H + HCO $\rightarrow$ H$_2$CO and H + HCO $\rightarrow$ CO + H from 0.5 each to 0.33 and 0.67, respectively (according to~\citealt{2021NatAs...5..197I}), in order to vary the amount of H$_2$CO and CO;
    \item \replaced[id=VS]{$E_a$}{E$_a$} of H$_2$CO + H $\rightarrow$ HCO + H$_2$ reaction from 2960 to 2470~K, to slow down the H$_2$CO production;
    \item \replaced[id=VS]{$E_a$}{E$_a$} of CH$_3$OH + H $\rightarrow$ CH$_3$O + H$_2$ and CH$_3$OH + H $\rightarrow$ CH$_2$OH  + H$_2$ from 6640 to 5530~K and from 4380 to 3610~K, respectively (according to~\citealt{2019A&A...627A...1Q}), to regulate methanol destruction. 
\end{enumerate}

All these changes give us an agreement between laboratory and modelling values with \replaced[id=VS]{RSS}{TSD} = 1.16, where the highest impact comes from the CO abundance. Considering the uncertainties associated with chemical desorption, this set gives us the best agreement between experimental and model results.  

Taking into account all the changes described above, we performed calculations for Experiments \#4-\#8, where the surface temperature was changed. Table~\ref{tab:MonacoDifTempExp4_8} shows the modelling results for all 5 molecules on the dust surface: CO, HCO, H$_2$CO, CH$_3$O and CH$_3$OH. The difference in abundances between models is not as high as in the case of laboratory experiments. However, modelling results are still in good agreement with them. In experiments with \replaced[id=VS]{$T_{dust}$}{T$_{dust}$} = 15~K we have \replaced[id=VS]{RSS}{TSD} = 0.2, which indicates a great agreement between the results.

\begin{table*}[h!]
\caption{Laboratory results (from Table~\ref{tab:experiments} and results of simulations with the MONACO code for $\{$CO + H$\}$ 60 min deposition \{CO+H\} with different surface temperatures (Experiments \#4-\#8)}             
\label{tab:MonacoDifTempExp4_8}    
\centering 
\begin{tabular}{c c c c c c c c}  
\hline\hline                 
Experiment & $T_{surf}$ & CO & HCO & H$_2$CO & CH$_2$OH & CH$_3$O & CH$_3$OH \\ 
& (K) & \multicolumn{6}{c}{(ML)}\\ 
\hline       
Laboratory results &&&&&&& \\ \hline
\#4 &   9 & 4.063 & - & 0.169 & - & - & 0.039\\      
 \#5 &   10 & 4.478 & - & 0.105 & - & - & 0.031\\
 \#6 &   11 & 5.948 & - & 0.053 & - & - & 0.021\\
 \#7 &   12 & 6.448 & - & 0.036 & - & - & 0.015\\
 \#8 &   15 & 6.593 & - & 0.018 & - & - & 0.010\\ \hline 
Modelling results &&&&&&& \\ \hline
\#4 &   9 & 5.437 & 1.260(-1) & 3.164(-1) & 4.796(-3) & 1.006(-2) & 3.758(-2) \\      
 \#5 &   10 &  5.545   & 3.900(-2) & 3.065(-1) & 1.454(-3) & 2.035(-3) & 3.242(-2) \\
 \#6 &   11 &  5.591   & 2.134(-2) & 3.050(-1) & 7.243(-4) & 9.834(-4) & 3.137(-2) \\
 \#7 &   12 &  5.64   & 1.363(-2) & 3.054(-1) & 4.524(-4) & 6.206(-4) & 3.102(-2) \\
 \#8 &   15 & 6.125   & 9.528(-4) & 2.894(-1) & 2.286(-5) & 3.926(-5) & 2.546(-2) \\ \hline                                  
\end{tabular}
\end{table*}

\subsection{Experiments \#9-\#10: {CO}+{H} experiments}\label{subsec:exp_9_10}

\begin{figure}[h!]
    \begin{minipage}{0.7\linewidth}
     \centering
     \includegraphics[width=\linewidth]{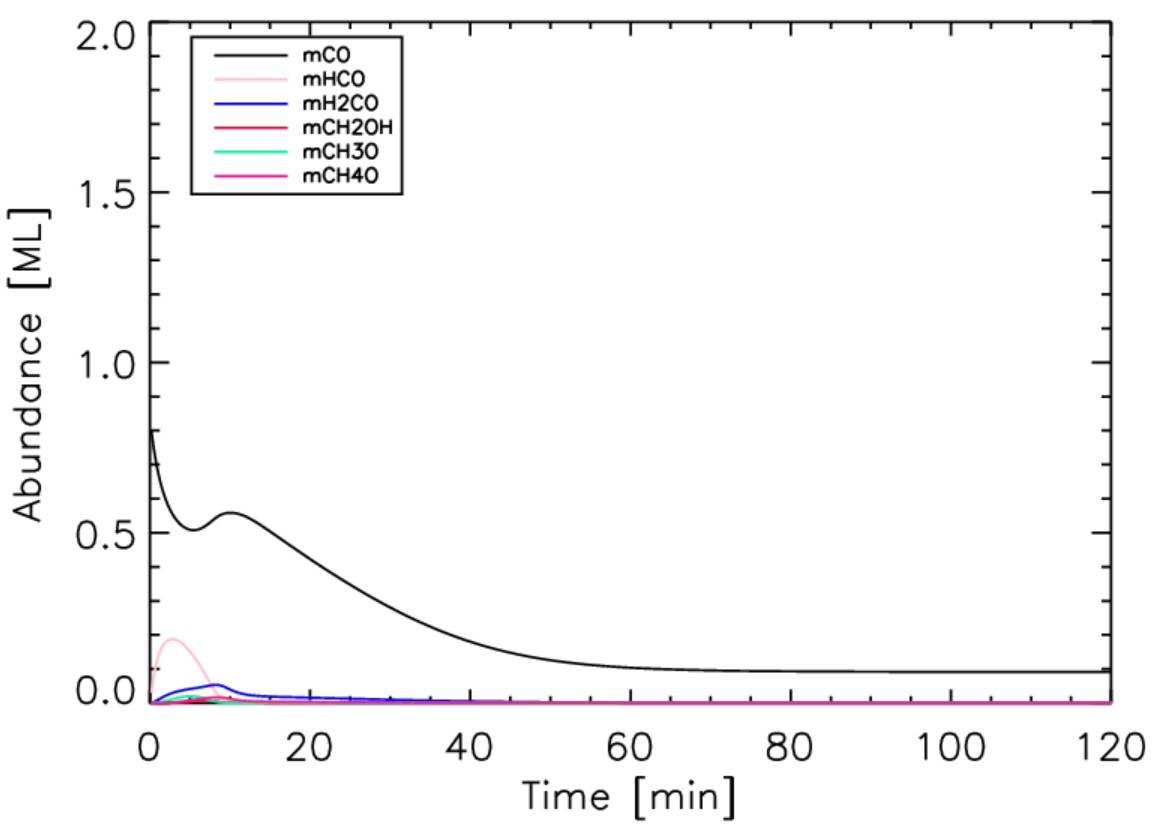}
     \end{minipage}\\
     \begin{minipage}{0.7\linewidth}
     \centering
     \includegraphics[width=\linewidth]{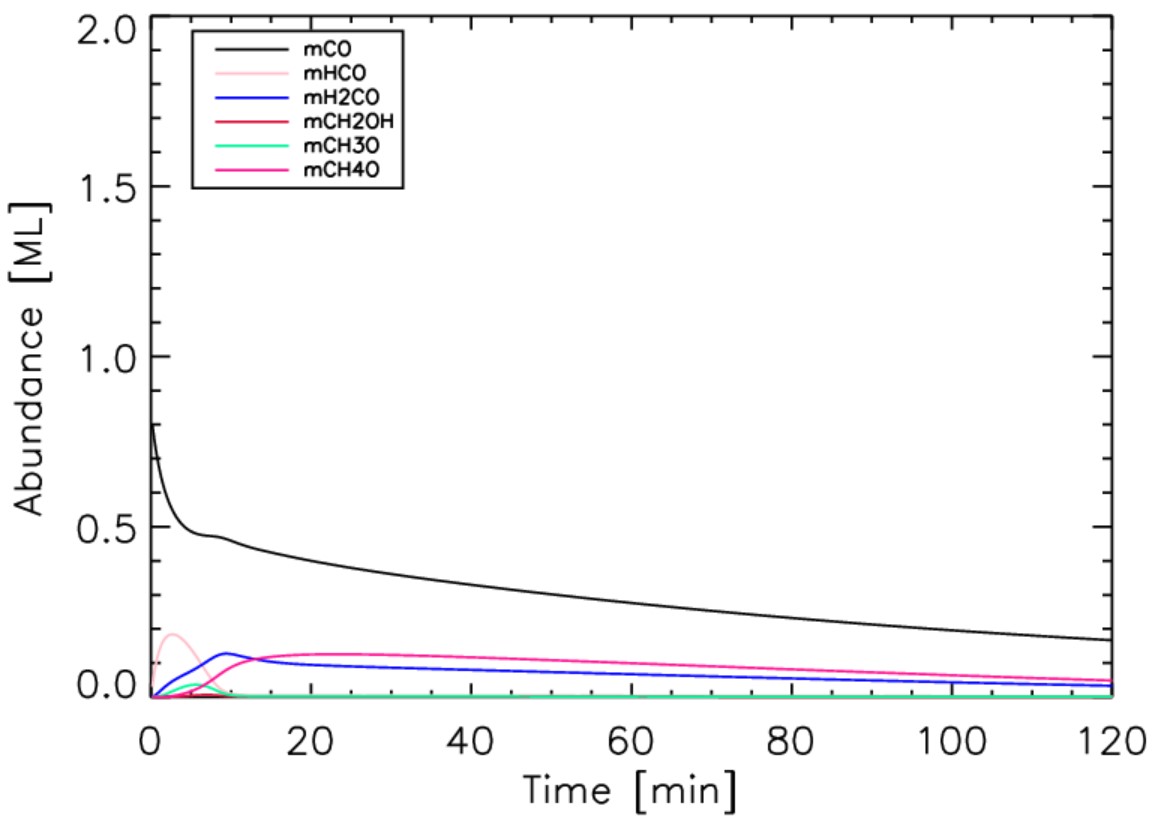}
     \end{minipage}     
    \caption{Modelling results for deposition of CO followed by deposition of H (Experiment \#9 with hydrogenation time of 120 min and \#10 with hydrogenation time of 75 min) with best-fit network (on the top) and standard network (on the bottom). mX (where X is the chemical formula of the compound) denotes molecules from bulk and surface together.}
    \label{fig:monaco_exp9_10}
\end{figure}

In this section, we show the results of modelling for experiments where the pre-deposited CO ice is hydrogenated. Figure~\ref{fig:monaco_exp9_10} shows the results of the modelling with the following initial parameters:

\begin{enumerate}
    \item $n = 5.3 \times 10^{7}~\text{cm}^{-3}$ - as gas number density ($n_H$);
    \item \replaced[id=VS]{$T_{gas}$}{T$_{gas}$} = 300~K, \replaced[id=VS]{$T_{d}$}{T$_{d}$} = 9~K (see Experiments \#8 -\#9 in Table~\ref{tab:experiments});
    \item initial abundances: \replaced[id=VS]{$n(H)$}{n(H)}/$n_H$ = 1, \replaced[id=VS]{$n(gCO)$}{n(gCO)}/$n_H$ = 8.53 $\times$ 10$^{-7}$; \replaced[id=VS]{$n(bCO)$}{n(bCO)}/$n_H$ = 6.42 $\times$ 10$^{-7}$, where abundances for gCO and bCO are taken from the modelling CO-only deposition for 9 minutes (according to the time of Experiments \#8 -\#9 at the Table~\ref{tab:experiments});
    \item two chemical networks: the one described above with the best-fit \replaced[id=VS]{$E_a$}{E$_a$}(CO+H) and \replaced[id=VS]{$E_b$}{E$_b$}(H) values, and the standard one from Table~\ref{tab:monaco_ch3oh}, where only $E_b$(H) = 500~K is changed.
\end{enumerate}

We note a non-monotonic and non-linear behaviour of CO in the first minutes (shown in both panels of Figure~\ref{fig:monaco_exp9_10}) as a result of the HCO destruction and transition back to CO. This raises the problem with IR detection of radicals (HCO, CH$_3$O or CH$_2$OH) that have never been detected during our experiments. For comparison, the decrease in the IR signal of CO is quite linear (without fast increases/decreases as seen in the bottom panel of Figure~\ref{fig:monaco_exp9_10}), and both H$_2$CO and CH$_3$OH are detectable. Thus, while modifying diffusion and reactivity barriers and adapting branching ratios can significantly improve results, there are still issues to be addressed, especially the too high accumulation of radicals predicted by the model.

\section{NAUTILUS code and results}\label{sec:nautilus}

Nautilus is a public astrochemical code that calculates chemical abundances (in gas-phase and on interstellar grain surfaces) as a function of time for given physical conditions. First presented in~\citealt{2016MNRAS.459.3756R}, this three-phase code can calculate not only the gas chemistry but also the chemistry of surface species in the outer monolayers (2 in the case of the present research) and the bulk mantle underneath.

The calculation of the chemical evolution of species is based on the rate equation method described above (Section~\nameref{sec:re}, \cite{1993MNRAS.263..589H}). The competition between diffusion, desorption, and reaction (as discussed by \citealt{2007A&A...469..973C} and \citealt{2011ApJ...735...15G}) is included in the model. In more detail, the NAUTILUS code is described in~\citealt{2016MNRAS.459.3756R}. All model parameters are explicitly provided in the code and can be adjusted for specific applications.

\begin{figure}
    \centering
    \includegraphics[width=0.7\linewidth]{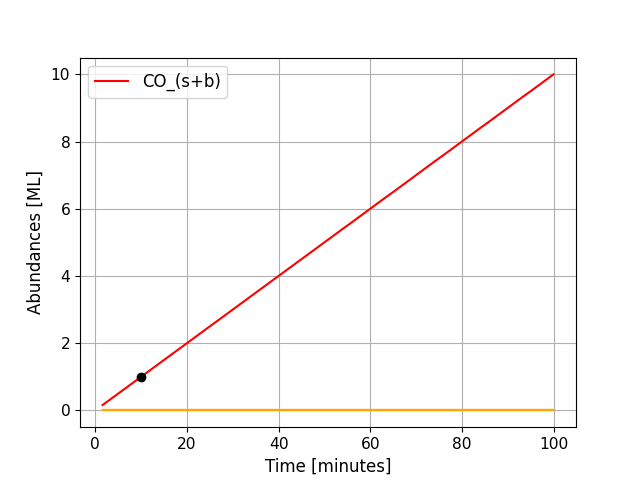}
    \includegraphics[width=0.7\linewidth]{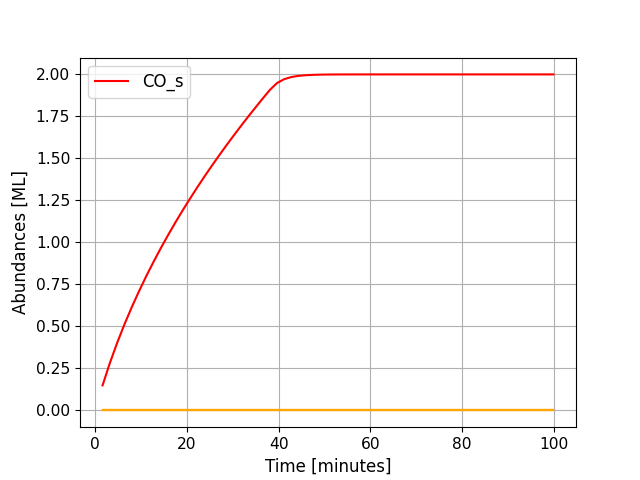}
    \caption{Experiment \#1: Top panel : Growth of the CO ice on the surface, using the NAUTILUS code. CO(s+b) denotes molecules from bulk and surface together. Bottom panel : Saturation of the chemical layer, the active layer is equal to 2. Black dot shows results from the experiments, where 1~ML of CO on the surface is obtained in 10 minutes.}
    \label{fig:naut_co_10min}
\end{figure}

We decided to calibrate fluxes and find the monolayer conversion coefficient using a mixture of H$_2$ and CO together, since these two molecules do not react. The reason for this choice is that the abundances are given in relative values to H, and we have retained this approach to avoid potential numerical problems. Figure~\ref{fig:naut_co_10min} represents the CO deposition of one ML in 10 minutes to mimic experiments listed in Table~\ref{tab:experiments}. Here, we found that with \replaced[id=VS]{$n(H)$}{n(H)} = $2.1 \times 10^{9}~\text{cm}^{-3}$ the modelling results match the results of Experiment~\#1 and give 1~ML of CO ice in 10 minutes (see Fig.~\ref{fig:naut_co_10min} for the linear growth of the CO monolayer and for the layer saturation). 
We observe a slightly different behaviour of the shape of the saturation curve of surface species compared to Figure~\ref{fig:monaco_co_only} using the MONACO code. This is not only due to the thickness of the chemical layer (2 vs\added[id=VS]{.} 1), but may also shed light on the different treatments of the surface to bulk conversion.

\begin{table*}[h!]
\caption{Results of simulations with the NAUTILUS code for $\{$CO + H$\}$ 60 min deposition with different surface temperatures (Experiments \#4-\#8). Tunnelling of H diffusion is turned on; diffusion barrier thickness is 2.5$\times 10^{-8}$ cm.}             
\label{tab:NautilusDifTempExp4_8_2}    
\centering  
\begin{tabular}{c c c c c c c c}  
\hline\hline                 
Experiment & T & CO & HCO & H$_2$CO & CH$_2$OH & CH$_3$O & CH$_3$OH \\ 
& (K) & \multicolumn{6}{c}{(ML)}\\ 
\hline                       
\#4 &      9 & 5.089  & 0.124 & 1.034 & 0.001  & 0.011 & 0.309 \\      
\#5 &      10 &  5.089   & 0.124 & 1.034 & 0.001  & 0.011 & 0.309 \\
 \#6 &     11 &  5.089   & 0.124 & 1.034 & 0.001  & 0.011 & 0.309 \\
\#7 &      12 &  5.089   & 0.124 & 1.034 & 0.030 & 0.011 & 0.309 \\
\#8 &      15 & 5.098   & 0.118 & 1.033 & 0.001 & 0.010 & 0.307 \\ 
\hline                                  
\end{tabular}
\end{table*}

\begin{table*}[h!]
\caption{Results of simulations with the NAUTILUS code for $\{$CO + H$\}$ 60 min deposition with different surface temperatures (Experiments \#4-\#8). Tunnelling diffusion of H is turned off. }             
\label{tab:NautilusDifTempExp4_8_1}    
\centering  
\begin{tabular}{c c c c c c c c}  
\hline\hline                 
Experiment & T & CO & HCO & H$_2$CO & CH$_2$OH & CH$_3$O & CH$_3$OH \\ 
& (K) & \multicolumn{6}{c}{(ML)}\\ 
\hline                       
 \#4 &     9 & 2.848 & 1.888 & 1.318 & 0.0576 & 0.277 & 0.203 \\      
\#5 &      10 &  2.863   & 1.856 & 1.329 & 0.057 & 0.255 & 0.230 \\
 \#6 &     11 &  3.070   & 1.643 & 1.306 & 0.049 & 0.240 & 0.279 \\
 \#7 &     12 &  3.662   & 1.105 & 1.226 & 0.030 & 0.178 & 0.382 \\
 \#8 &     15 & 5.098   & 0.118 & 1.033 & 0.001 & 0.011 & 0.307 \\ 
\hline                                  
\end{tabular}
\end{table*}

\begin{table*}[h!]
\caption{Results of simulations with the NAUTILUS code for $\{$CO + H$\}$ 60 min deposition with different surface temperatures (Experiments \#4-\#8). Tunnelling of H diffusion is turned on; the diffusion barrier thickness is 3.5$\times 10^{-8}$ cm. }             
\label{tab:NautilusDifTempExp4_8_3}    
\centering  
\begin{tabular}{c c c c c c c c}  
\hline\hline                 
Experiment & T & CO & HCO & H$_2$CO & CH$_2$OH & CH$_3$O & CH$_3$OH \\ 
& (K) & \multicolumn{6}{c}{(ML)}\\ 
\hline                       
\#4 &      9 & 2.964   & 1.750 & 1.319 & 0.053   & 0.247 & 0.256  \\      
\#5 &      10 &  2.964   & 1.750 & 1.319 & 0.053   & 0.247 & 0.256  \\
 \#6 &     11 &  3.070   & 1.643 & 1.306 & 0.049   & 0.240 & 0.279  \\
\#7 &      12 &  3.662   & 1.105 & 1.226 & 0.030 & 0.178  & 0.382 \\
\#8 &      15 & 5.098   & 0.118 & 1.033 & 0.001 & 0.011 & 0.307 \\ 
\hline                                  
\end{tabular}
\end{table*}

Following the same scenario for MONACO and pyRate codes, we have simulated the different experiments one after another, starting with Experiment \#4. Table \ref{tab:NautilusDifTempExp4_8_2} shows the results obtained with the fiducial NAUTILUS input parameters, including the hydrogen diffusion parameters.
We find the same general trends: 
\begin{itemize}
    \item almost no temperature dependency;
    \item a moderate overestimation of the molecules on the surface, indicating low efficiency of the chemical desorption process (independent of the H flux), compared to the experimental efficiency;
    \item an overabundance of radical intermediates.
\end{itemize}

In the Nautilus code, the hydrogen diffusion treatment can be set to only thermal hopping or both tunneling and hopping. The width of the barrier diffusion for H can be changed so that the ratio between thermal hopping and tunneling can be adapted.

Tables \ref{tab:NautilusDifTempExp4_8_2}, and \ref{tab:NautilusDifTempExp4_8_3} show the results of changing the diffusion parameters of H. In the first case, when the diffusion is only thermal, a temperature dependence is observed. In the second case, we keep the tunneling diffusion, but enlarge the diffusion barrier width to 3.5 {\AA}, so that we increase the crossover temperature. As expected, the model results are closer to the thermal diffusion case only, but with some modifications at the lowest temperatures.

However, while these tests illustrate the great importance of H diffusion, they nevertheless show that it is not enough to modify these parameters to resolve the entire successive chemical chain, which depends above all on the number density of H on the surface, and therefore on its diffusion. This is somewhat counterintuitive, but the more H diffuses, the more it reforms H$_2$, and the lower its number density, and therefore the less effective the successive hydrogenation of CO, which has an entry barrier. We have also performed a full minimization of the reaction barrier parameters and found the best values, which are compatible with the values constrained in the previous section using the MONACO code.

\section{pyRate code and results} \label{sec:pyrate}

pyRate is a rate-equation based astrochemical code that simulates the gas-grain chemical interaction occurring through adsorption and several desorption mechanisms. As described in detail in \cite{sipila2015benchmarking}, pyRate follows the time evolution of the abundances and reactions of chemical species with chemical networks developed to track the spin states of several species (\ce{H2}, \ce{H2+}, \ce{H3+}, \ce{NH3}, and \ce{H2O}; \citealt{sipila2013hd,sipila2015benchmarking,sipila2015spin}). Grain-size distributions can also be considered \cite{sipila2020effect}. Like MONACO and Nautilus, the three-phase model (gas/surface/mantle) is considered for the experiment by co-deposition (experiments \#1-8), where the reactive surface is defined at one layer but with no diffusion and reactivity in the mantle \cite{1993MNRAS.263..589H}. The two-phase model (gas/solid) is used for the hydrogenation of pre-deposited CO ice experiment (experiment \#9 and \#10) because the abundance of species in the solid-phase is less than one monolayer.

For the following calculation, we use the same physical parameters as those described previously in Section\,\nameref{sec:re}, with some differences. The hydrogen number density $n_{\ce{H}}$ is set at $5\times10^{10}$ cm$^{-3}$. The probability of a reaction via diffusion is set at one for radical-radical reactions, or can be calculated by tunnelling effect, with the chemical network summarised in Table\,\ref{tab:monaco_ch3oh} (the column $E_PR$ summarizes energies used in the code). The width of the tunnelling barrier for diffusion and reaction is assumed to be one. Reaction-diffusion-competition can be included or ignored. Chemical desorption is estimated using Riedel's equation\cite{riedel2023modelling}, a modified form of equation\,\ref{eq: chemical desorption}, in which the excess energy from the reaction is distributed among the products according to their mass.

\begin{figure}[h!]
    \centering
    \includegraphics[width=0.7\columnwidth]{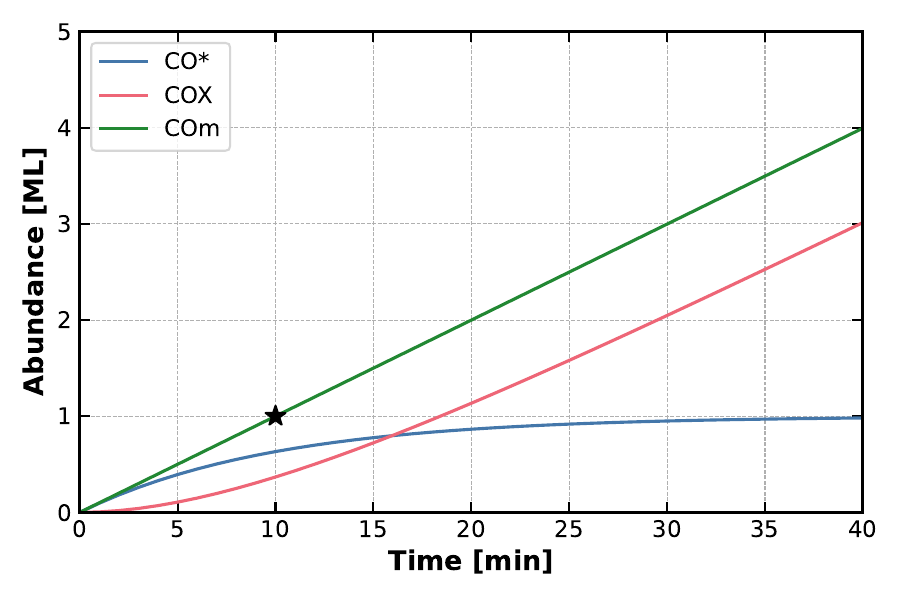}
    \caption{Experiment \#1: Growth of CO ice on the substrate, case of pyRate code. The blue curve designated as CO* illustrates the CO molecules located on the surface, while the red curve labeled COX indicates the CO molecules present within the bulk. The green curve, labeled COm, represents the sum of the CO molecule present on the surface and in the bulk. \added[id=BH]{The black star represents results from the experiments \#1.}}
    \label{fig pyRates: pyRate_COonly}
\end{figure}

In the case of the pyRate code, beam flux calibration was performed by measuring the time needed for a monolayer of the species to be completed on the grain surface, which was done by adjusting the initial abundances accordingly. All possible surface reactions were turned off during this calibration. The initial abundances were adjusted until they reached $n(\ce{CO})/n_{\ce{H}}=4.203\times10^{-3}$ and $n(\ce{CO})/n_{\ce{H}}=4.669\times10^{-3}$ for CO in experiments \#1-2 and \#3-10, respectively. These values correspond to deposition times of 10 minutes for a monolayer in the first set (see Figure\,\ref{fig pyRates: pyRate_COonly}) and $9$\,minutes in the second. For atomic hydrogen, the initial abundances were $n(\ce{H})/n_{\ce{H}}=7.676\times10^{-3}$ for experiments \#1-2 and $n(\ce{H})/n_{\ce{H}}=3.839\times10^{-3}$ for experiments \#3-10, considering a sticking coefficient of 0.41 \cite{cuppen2010h2}. The sticking coefficient of CO is around 1 in the experimental conditions \cite{he2016sticking}.

\begin{table*}[h!]
    \caption{pyRate results with tunnelling diffusion and no reaction-diffusion competition.}
    \label{tabA: result pyRate no competition}
    \centering
    \begin{adjustbox}{width=1\textwidth}
    \begin{tabular}{*{10}{c}}
    \hline
    \# & T & Time &   CO & \ce{H2CO} & \ce{CH3OH} & Missing & HCO & \ce{CH2OH} & \ce{CH3O} \\
      & (K) & (minutes) & \multicolumn{7}{c}{(ML)} \\
    \hline
    1 & 10 & 15 & 1.500(+0) & & & & & & \\
    2 & 10 & 15 & 1.433(+0) & 2.788(-2) & 5.729(-5) & 3.860(-2) & 7.666(-8) & 1.173(-12) & 1.617(-10)\\
    \hline
    3 & 15 & 60 & 6.667(+0) & & &  & & & \\
    4 & 9 & 60 & 6.329(+0) & 1.403(-1) & 3.864(-4) & 1.976(-1) & 3.388(-7) & 7.907(-12) & 8.135(-10)  \\
    5 & 10 & 60 & 6.328(+0) & 1.403(-1) & 3.862(-4) & 1.979(-1) & 3.388(-7) & 7.904(-12) & 8.133(-10) \\
    6 & 11 & 60 & 6.328(+0) & 1.402(-1) & 3.860(-4) & 1.983(-1) & 3.388(-7) & 7.900(-12) & 8.131(-10) \\
    7 & 12 & 60 & 6.328(+0) & 1.402(-1) & 3.859(-4) & 1.988(-1) & 3.388(-7) & 7.896(-12) & 8.129(-10) \\
    8 & 15 & 60 & 6.326(+0) & 1.401(-1) & 3.853(-4) & 2.008(-1) & 3.387(-7) & 7.886(-12) & 8.123(-10) \\
    \hline
    9 & 9 & 9+120 & 4.574(-1) & 2.271(-1) & 5.426(-3) & 3.101(-1) & 2.463(-8) & 1.108(-10) & 1.319(-9) \\
    10 & 9 & 9+75 & 6.128(-1) & 1.649(-1) & 2.390(-3) & 2.200(-1) & 3.291(-8) & 4.883(-11) & 9.567(-10)\\

    \hline
    \end{tabular}
    \end{adjustbox}
\end{table*}

\begin{figure}[h!]
    \centering
    \includegraphics[width=0.7\columnwidth]{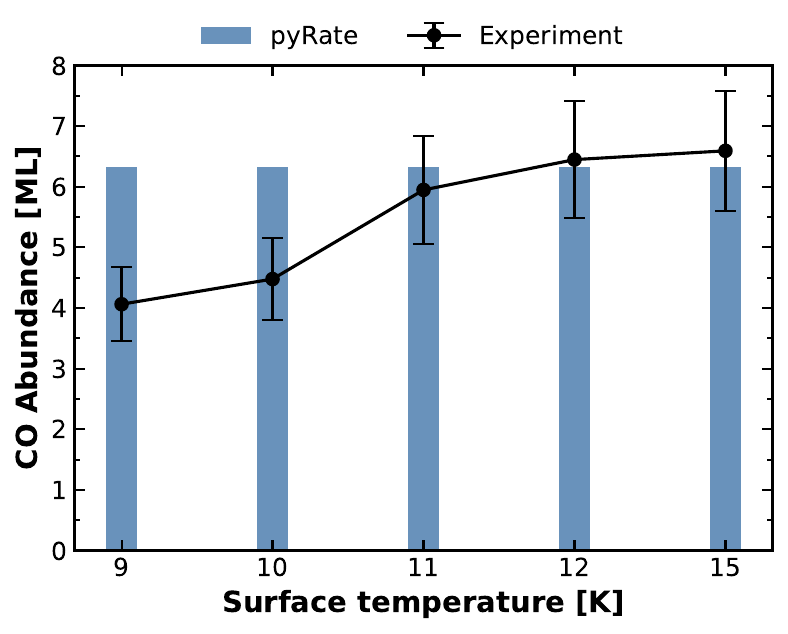}
    \caption{Solid-phase \ce{CO} abundance as a function of surface temperature after 60\,minutes of \{\ce{CO + H}\} co-deposition (experiments \#4-8). The blue bar shows the predictions from the pyRate simulation (Table\,\ref{tabA: result pyRate no competition}), while the black dots connected by lines indicate the experimental result (Table\,\ref{tab:experiments}).}
    \label{fig pyRates: figure_CO_Tdependence}
\end{figure}

The results obtained with pyRate for experiments \#1 to \#10 are presented in Table\,\ref{tabA: result pyRate no competition} with tunnelling diffusion and without reaction-diffusion competition. Figure\,\ref{fig pyRates: figure_CO_Tdependence} illustrates the \ce{CO} abundance in the solid phase as a function of surface temperature during \ce{CO} and \ce{H} co-deposition experiments (\#4-8). The pyRate predictions (represented by the blue bar) suggest a temperature-independent \ce{CO} abundance in the solid phase of $\sim 6.329$\,ML between $9$ and $15$\,K. In contrast, the experimental measurements (black dots connected by lines) reveal a clear temperature dependence, with \ce{CO} abundances of $4.063$\,ML at $9$\,K and $6.593$\,ML at $15$\,K. This discrepancy results from how the model handles reactivity and diffusion processes, which were both calculated with the tunnelling effect. Tunnelling in chemical reactions depends primarily on the mass of the involved species, and on the width and energy of the potential barrier. For the \ce{CO + H -> HCO} reaction, \citealt{andersson2011tunneling} has demonstrated theoretically that tunnelling dominates at temperatures below $181.8$\,K in the gas-phase. In addition, \citealt{2014A&A...572A..70R} has shown theoretically that in the solid phase, within clusters of water molecules, the crossover temperature is $141$\,K for \ce{CO + H -> HCO} and $181$\,K for \ce{H2CO + H -> CH3O}, which allows us to conclude that the tunnelling effect dominates the entire \ce{CO + H} chemical network in our experimental conditions.
This is why we attribute the temperature dependence to thermal diffusion coupled with competitive reaction-diffusion processes. Thermal diffusion, which follows an Arrhenius law, increases exponentially with temperature. Conversely, the competitive reaction-diffusion mechanism, proposed by \citealt{2011ApJ...735...15G}, suggests that if the migration time is shorter than the reaction time, there is a non-zero probability that a reactant may migrate to a different site instead of reacting with another species. Therefore, the higher the temperature, the faster \ce{H} atoms diffuse and the less they react with \ce{CO}. The probability for species A and B to react and form a species AB is described as follows:
\begin{equation} \label{eq: competition reaction-diffusion}
    p_{\text{AB}} =\left\{  
    \begin{array}{ll}
         \kappa_{\text{AB}} & \text{no competition,}  \\
         
        \frac{\nu_{\text{AB}} \kappa_{\text{AB}}}{\nu_{\text{AB}} \kappa_{\text{AB}} + k_{\text{diff}}(\text{A})+ k_{\text{diff}}(\text{B})} & \text{with competition,}
    \end{array}
     \right.
\end{equation}
where $\nu_{\text{AB}}$ is the highest characteristic frequency between reactant A and B, $\kappa_{\text{AB}}$ the reaction probability from \citealt{hasegawa1992models}, and $k_{\text{diff}}$ the diffusion rate coefficient. The results from MONACO and NAUTILUS (Tables\,\ref{tab:MonacoDifTempExp4_8}\,and\,\ref{tab:NautilusDifTempExp4_8_1}) show that the reaction-diffusion competition coupled with thermal hopping diffusion follows the temperature dependence of CO abundance observed in the experimental data (Table\,\ref{tab:experiments}). However, this approach also presents additional issues.

\begin{figure}[h!]
    \centering
    \includegraphics[width=0.7\columnwidth]{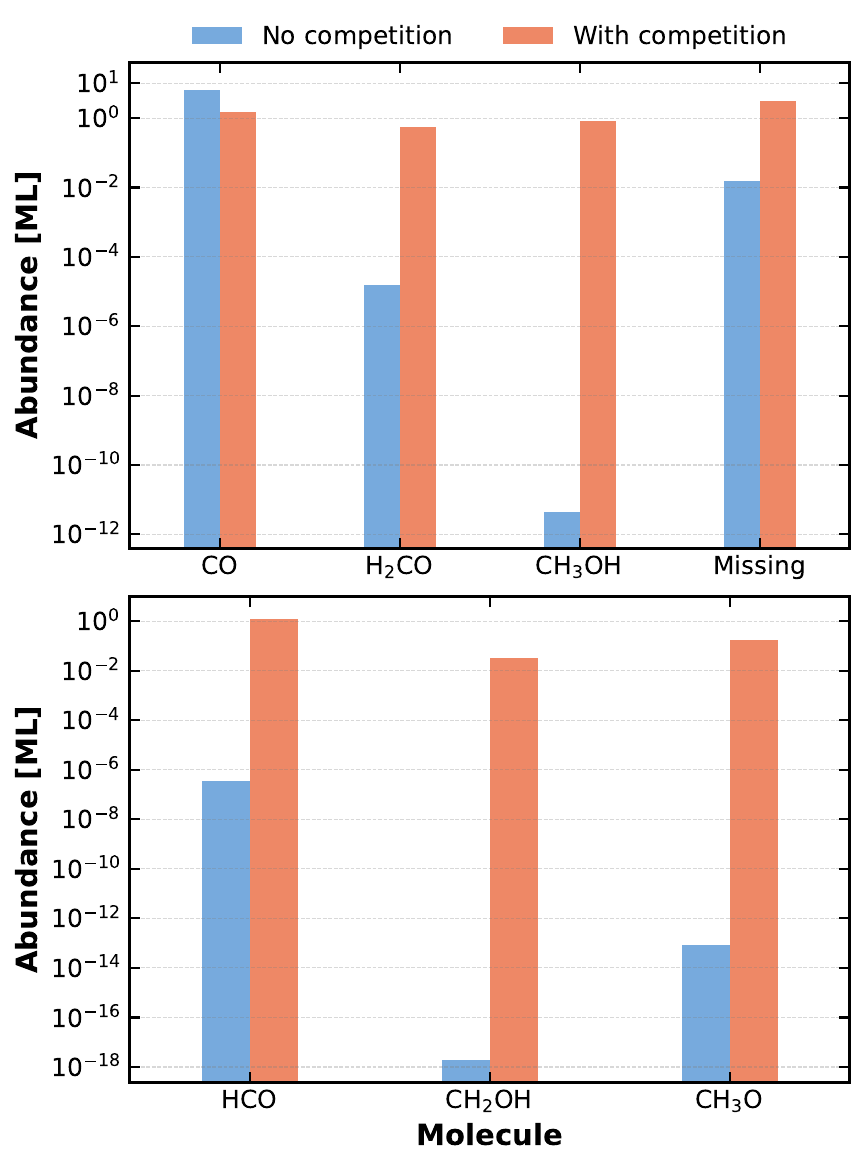}
    \caption{Results of pyRate simulations for \{\ce{CO + H}\} co-deposition at a surface temperature of 10\,K (experiment \#5) with thermal hopping diffusion. The data are presented for scenarios without reaction-diffusion competition (blue bar) and with reaction-diffusion competition (orange bar). The top panel shows the solid-phase abundances of detectable molecules (\ce{CO}, \ce{H2CO}, and \ce{CH3OH}), where ''Missing'' indicates molecules absent from the solid-phase due to chemical desorption. The bottom panel illustrates the solid-phase abundances of radicals produced through \ce{CO} reacting with \ce{H}, especially \ce{HCO}, \ce{CH2OH}, and \ce{CH3O}, which are not detected in the experiment.}
    \label{fig pyRates: competition_plot}
\end{figure}

Figure\,\ref{fig pyRates: competition_plot} presents the pyRate results for experiment \#5, comparing scenarios with no reaction-diffusion competition (blue) and with competition (orange), while maintaining the same physical parameters. Thermal hopping diffusion is also considered. The results indicate that \ce{H} atoms exhibit increased reactivity with \ce{CO} molecules under diffusion-reaction competition conditions, with $1.525$\,ML of \ce{CO} remaining on the solid-phase at the end of the co-deposition \{\ce{CO + H}\}, compared to $6.352$\,ML in its absence. This trend is also observed for other products: $0.5569$\,ML of \ce{H2CO}, $0.8157$\,ML of \ce{CH3OH}, and $3.038$\,ML for the missing column (chemical desorption) when competition is considered, compared to $1.541\times10^{-5}$\,ML, $4.459\times10^{-12}$\,ML, and $1.487\times10^{-2}$\,ML without competition, respectively. Regarding radicals, a notable difference in their estimated abundances is observed when employing the competition method, with \ce{HCO} reaching up to $1.257$\,ML. Given this abundance, \ce{HCO} should be detectable through IR spectroscopy during the experiments, however, such detection is not observed (see Figure\,\ref{fig_exp: IR spectra and radicals}).

A closer look at equation\,\ref{eq: competition reaction-diffusion} for the \ce{CO + H -> HCO} reaction reveals that the reaction coefficient rate $\nu_{\text{AB}}\kappa_{\text{AB}}$, estimated approximately at $1.8\times10^5$\,$s^{-1}$ for a barrier of $1730$\,K, is much higher than the diffusion coefficient rates $k_{\text{diff}}$ of H atoms and \ce{CO} molecules, estimated approximately at $6.0\times10^1$ and $2.8\times10^{-21}$\,s$^{-1}$, respectively. With a reaction coefficient that is approximately four orders of magnitude greater than hydrogen diffusion, equation\,\ref{eq: competition reaction-diffusion} predicts a probability near unity when the reaction-diffusion competition is used. Thus, the inclusion of the diffusion-reaction competition makes the reaction ''barrierless'' and leads to an over-reactivity of \ce{CO} with \ce{H}. Therefore, all \ce{H} atoms mainly react with \ce{CO} and less frequently with other molecules, since \ce{CO} is present everywhere on the surface as co-deposited with \ce{H}.

Reaction-diffusion competition currently does not address the experimental question. It effectively reproduces the temperature dependence, suggesting an interaction between reactivity and surface processes, mainly diffusion at low temperatures. However, in regimes where diffusion is slower than reactivity, the model tends to overestimate the reaction probability.

\section{Conclusion}\label{sec:comparison}

This study has shown that rate-equation codes \replaced[id=VS]{such as}{like} MONACO, NAUTILUS, and pyRate can be used to simulate experimental conditions without the need \replaced[id=VS]{to change}{of change of} the programming structure of dedicated computer codes. It simply requires a good understanding and proper adjustment of the physical parameters \added[id=VS]{in models}, and for experimentalists \added[id=VS]{---} a clear description of the experimental parameters and observables.
At 10\,K, adjusting the H diffusion property in any model is enough to obtain a fair reproduction of \added[id=VS]{the} experimental data. However, \replaced[id=VS]{our attempts to reproduce the}{when trying to reproduce} temperature dependency \added[id=VS]{showed that} the quantum tunneling crossover temperature appears to be lower or the quantum diffusion of H atoms at 10\,K is overestimated. Using thermal hopping for H atoms only allowed us to adjust reaction parameters to finally achieve a reasonably good match of the experiments, with two notable deviations: the underestimation of the chemical desorption and the overabundance of radicals proposed in the models. To accurately reproduce the surface temperature dependence, reaction-diffusion competition and thermal hopping diffusion must be incorporated. However, reaction-diffusion-competition results in over-reactivity when the diffusion coefficient rate is lower than the reaction coefficient rate, leading to an excess of radicals.

This exploratory work shows us that both modellers and experimentalists can compare their approaches, providing the former with a robust model and the latter with unprecedented access to the details of the processes at work in the experiments. 

After 18 months of parallel and complementary work, some points stand out very clearly. For experiments, in addition to an unambiguous description, it is crucial to have a coherent quantitative dataset, as complete as possible, not only on observables (deposited species, formed species including upper limits for radicals), but also on experimental parameters. In this study, we found that the surface temperature is very important, but the fluxes could also add valuable constraints for modellers. It would also be good to probe different chemical regimes (barrierless or not) and to propose layered experiments to test the molecule transfer from surface to bulk and vice versa, which are yet to be compared with models.

Before examining the details of chemical networks and their barriers, it seems that the key point lies in the diffusion of hydrogen and its intertwining with certain model assumptions, particularly the diffusion-reaction competition. In the case of hydrogenation reactions, these are an essential step in validation. To this end, experiments with different entrance barriers would impose different constraints and would undoubtedly help to tackle this issue. Adjusting reaction efficiencies does not seem to be a problem if the scope of conditions is large enough. It is likely that the importance of Eley-Rideal processes, for example, can also be addressed, since it is easy to vary the initial coverage and to test the hypothesis. 

\begin{acknowledgement}

This project was supported by the Programme National “Physique et Chimie du Milieu Interstellaire” (PCMI) of CNRS/INSU with INC/INP co-funded by Commissariat \`{a} l'Energie Atomique (CEA) and Centre National d Etudes Spatiales (CNES). This work was supported by the Agence Nationale de la Recherche (ANR) SIRC project (Grant ANR-SPV202448 2020-2024). AV work was supported by the contract FEUZ-2025-0003. The authors would like to thank Julie Vitorino and Saoud Baouche for their help with the experiments conducted for this project at the CY LIRA laboratory.

\end{acknowledgement}

\begin{suppinfo} \label{app: Additional Data}

\end{suppinfo}

\bibliography{achemso-demo}

\end{document}